\documentclass[aps,prb,twocolumn,groupedaddress,showpacs]{revtex4}
\usepackage{graphicx}

\begin{document}

\title{Nucleation of superconductivity in a mesoscopic loop of
varying width}

\author{Mathieu Morelle}
\email[]{mathieu.morelle@fys.kuleuven.ac.be} \homepage[]{http://www.fys.kuleuven.ac.be/vsm} \affiliation{Nanoscale Superconductivity and Magnetism Group,
Laboratory for Solid State Physics and Magnetism, K.U. Leuven, Celestijnenlaan 200D, B-3001 Leuven, Belgium}
\author{Dusan S. Golubovi\'c}
\author{Victor V. Moshchalkov}
\affiliation{Nanoscale Superconductivity and Magnetism Group, Laboratory for Solid State Physics and Magnetism, K.U. Leuven, Celestijnenlaan 200D, B-3001
Leuven, Belgium}
\date{\today}

\begin{abstract}
We study the evolution of the superconducting state in a perforated disk by varying the size of the hole. The superconducting properties are investigated
by means of transport measurements around the superconducting/normal phase boundary $T_c(H)$. A transition from a one-dimensional to a two-dimensional
regime is seen when increasing the magnetic field for disks with small holes. A good agreement is found between the measured $T_c(H)$ line and the
calculations performed in the framework of the linearized Ginzburg-Landau theory. The effect of breaking the axial symmetry of the structure by moving
the hole away from the center of the disk is also studied. An enhanced critical field is found for the asymmetric structures, possibly due to the
recovery of the singly connected state.
\end{abstract}


\maketitle
\section{Introduction}

The fluxoid quantization constraint in superconducting loops structures gives rise to the well-known periodic Little-Parks\cite{lp62orig} oscillations in
the $T_c(H)$ phase line. The confinement of the superconducting condensate in mesoscopic structures leads to interesting new phenomena that are strongly
dependent on the geometry and the topology of the structure\cite{vvm95nature}.

While the limiting cases of a superconducting disk\cite{sj65cyl,fink66presson,vvm97pme,deo97prl,benoist97zb,schweigert98disks} or a thin
loop\cite{lp62orig, vvm95nature,Liu01,berger95prl, berger97prb, bergerbook, vodolazov02transition, buzdin02} have been broadly studied experimentally and
theoretically, the intermediate case is not. The latter is also closely related to the problem of a thin film exposed to a parallel magnetic
field\cite{sj63lin,golubov90}. In this situation, a quasi one-dimensional (1D) behavior, characterized by a parabolic dependence of the $T_c(H)$ phase
line, was predicted by Saint-James and de~Gennes\cite{sj63lin} as long as the thickness $\tau$ (in units of $\xi(T)$) is smaller than approximately two.
Above this value, the two surface superconducting sheaths are separated by a normal region, and a linear $T_c(H)$ dependence is observed, typical for a
two-dimensional (2D) system in a perpendicular field. At this dimensional crossover point ($\tau$=1.84), vortices start to nucleate in the
sample\cite{schultens70zp, fink69crittick}.

In their pioneering experiment, Little and Parks\cite{lp62orig} studied the field-temperature $H-T$ phase diagram of a thin-wire loop in an axial
magnetic field. A periodic component in the experimental $T_c(H)$ line was found. In the framework of the nonlinear Ginzburg-Landau (GL) theory, Berger
and Rubinstein\cite{berger95prl, berger97prb} studied mesoscopic superconducting loops.  They have predicted that, if the thickness of the loop is not
exactly uniform, then there exist situations for which superconductivity is suppressed at a certain location, so that the superconducting loop
effectively becomes singly connected and no supercurrent² flows. When this happens, the sample is in the so called \emph{`singly connected state'}.
However, Vodolazov \emph{et al.} showed that the state where the order parameter vanishes at one point is still doubly connected since the phase of the
order parameter is not independent on both sides of the place where the order parameter is zero. They suggested to call this state a one-dimensional
vortex state.

The intermediate case of finite width loops was studied within the London theory by Bardeen\cite{bardeen61}. He calculated that the flux is quantized in
units of $\nu \Phi_0$ (with $\nu<1$) in cylinders of very small diameters and with a wall thickness of the order of the penetration depth. It was later
calculated that the flux through an area $S=\pi r_m^2$ is quantized in units of $\Phi_0$, with $r_m=(r_i+r_o)/2$ the arithmetic mean of the inner ($r_i$)
and outer ($r_o$) radius\cite{groff68}. Arutunyan and Zharkov\cite{arutyunyan83jltp} determined in the London limit an effective radius of
$r_{eff}=\sqrt{r_ir_o}$ such that inside this ring, the flux was exactly quantized. These two different values $r_m$ and $r_{eff}$ are nearly identical
for the narrow ring. Baelus~{\it et al.\/}\cite{baelus00} found that the value of $r_i<r_{eff}<r_o$ was dependent on the vorticity $L$ and in fact
$r_{eff}$ turns out to be an oscillating function of the magnetic field.

A self-consistent treatment of the full nonlinear GL equations for a square loop has been carried out by Fomin~{\it et
al.\/}\cite{fomin97ssc,fomin98prb}. The order parameter $|\Psi|$ distribution was found to be strongly inhomogeneous due to the presence of sharp
corners. The precise shape of the $T_c(H)$ curve crucially depends on the area fraction for which $T_c(H)\neq 0$.

Bruyndoncx~{\it et al.\/}\cite{bruyndoncx99crossover} investigated infinitely thin loops of finite width using the linearized GL equation. The induced
magnetic field can be neglected in this case. They limited their investigations to circular symmetric solutions so that only the giant vortex state was
studied. Berger and Rubinstein\cite{berger99prb} also considered infinitely thin loops of finite width using the nonlinear GL theory, neglecting the
induced fields.

Baelus~{\it et al.\/}\cite{baelus00} analyzed circular flat disks of nonzero thickness with a circular hole in it. They also investigated the case where
the hole is shifted off the center of the disk. The superconducting properties were studied also deep in the superconducting state. For small
superconducting disks with a hole in the center, they found only the giant vortex state. The influence of the radius of the hole on the superconducting
state was considered. For larger holes in perforated superconducting disks, a re-entrant behavior was seen, where a transition from the giant vortex
state to a state with separated vortices and back to the giant vortex state was found.

Recently, Pedersen~{\it et al.\/}\cite{pedersen01} investigated experimentally the magnetization of a mesoscopic loop. The periodicity observed in the
magnetization measurements reveals a sub flux quantum shift. This fine structure was interpreted as a consequence of a giant vortex state nucleating
towards either the inner or the outer side of the loop.

This experiment has lead to a recent growing interest in the mechanism of flux transition in superconducting
loops\cite{vodolazov02transition,vodolazov02phaseslip,berger03}. Multiple flux jumps and irreversible behavior of the magnetization were observed in thin
mesoscopic rings by Vodolazov~{\it et al.\/}\cite{vodolazov03}. At low magnetic field and for rings with sufficiently large radii, they showed
experimentally and theoretically, using the time-dependent GL theory, that the vorticity may change by values larger than one.

The existence of a zero-current line in mesoscopic superconducting rings has been found both theoretically\cite{berger95prl} and through experimental
observation of self-generated weak links\cite{vvm93nature}. It was suggested in Ref.~\onlinecite{dubonos03} that a system of asymmetric loops can be used
as noise detector or as source of high frequency radiation.

Beyond the vortex and the giant vortex configurations, the ring-like vortex solutions of the GL equations in superconducting mesoscopic devices were
investigated\cite{stenuit00,govaerts00,yampolskii00}. Those solutions possess a unique winding number in the whole ring, but the order parameter vanishes
on one or more cylindric surface. For a nanosized Pb bridge, it has been reported that the vorticity varies along the axis of the bridge\cite{misko01}.
Solving self-consistently the nonlinear GL equations for a mesoscopic superconducting ring, Zhao~{\it et al.\/}\cite{zhao03} obtained solutions with
different vorticity inside and outside the zero-current line at a certain radius. They, however did not consider the phase coupling of the order
parameter between the two superconducting parts of the ring separated by the zero-current line.

The paramagnetic response for a stable configuration of a mesoscopic ring has been studied in Ref.~\onlinecite{meyers03}. They found an oscillation of
the order parameter density profile when changing the vorticity.

Using ultrasensitive susceptibility techniques and scanning Hall probe microscopy Davidovi\'{c}~\emph{et al.}~\cite{davidovic96prl,davidovic97prb} have
studied arrays of electrically isolated superconducting mesoscopic rings. When these rings are biased in an external flux of $\Phi_0/2$, they can be in
either of two energetically degenerated fluxoid states. The magnetic moments produced by the supercurrents in these rings are analogous to Ising spins,
and neighboring rings interact antiferromagnetically via their dipolar magnetic fields. The ring dynamics is dominated by an energy barrier between the
two states which rises rapidly as the temperature is lowered below $T_c$.

In this paper we shall study the systematic variation of the superconducting phase  boundary, $T_c(H)$, in perforated disks with different $r_i/r_o$
ratios, which realize a cross-over from the singly-connected disk to the limit of the thin ring. The rest of the paper is arranged as follows: in
Section~\ref{sec:rings}, we will study the evolution of the superconducting state for the transition from a disk geometry to a thin ring. The
superconducting properties of the disks with a hole in the center will be analyzed by transport measurements carried out around the
superconducting/normal transition line. In Section~\ref{sec:nonsym}, the effect of breaking the axial symmetry of the structure by shifting the hole off
the center of the disk will be discussed. The onset of dissipation below $T_c(H)$ will be studied in Section~\ref{sec:diss}.

\section{Superconducting rings}
\label{sec:rings}
\subsection{Sample properties}

\begin{figure}
\centering
\includegraphics*[width=8.5cm,clip=]{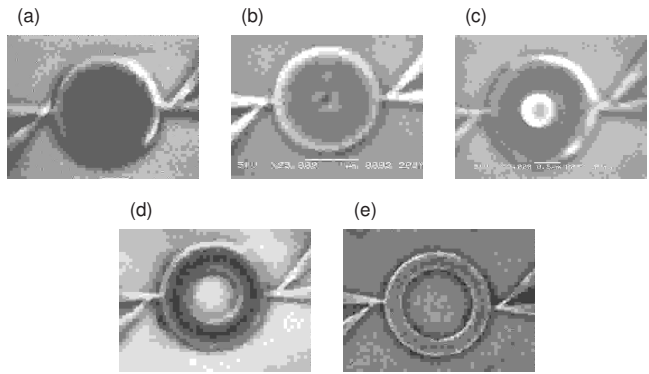}
\caption{SEM micrograph of (a) an Al disk with outer radius $r_o$=1~$\mu m$ and of a loop with outer radius $r_o$=1~$\mu m$ and inner radius (b)
$r_i$=0.1~$\mu m$, (c) $r_i$=0.3~$\mu m$, (d) $r_i$=0.5~$\mu m$ and (e) $r_i$=0.7~$\mu m$.} \label{Fig:SEMLoops}
\end{figure}

A SEM micrograph of the different studied samples prepared with e-beam lithography is given in Fig.~\ref{Fig:SEMLoops}. All the structures consist of
disks with external radii of $r_o=1~\mu m$. The radii of the holes (Fig.~\ref{Fig:SEMLoops}, determined from SEM micrograph, were $r_i=0~\mu m$ (a),
$r_i=0.1~\mu m$ (b), $r_i=0.3~\mu m$ (c), $r_i=0.5~\mu m$ (d) and $r_i=0.7~\mu m$ (e). All the samples were evaporated in the same run, except for the
thinnest loop. A different evaporation will only slightly alter the superconducting properties like the coherence length and the critical temperature.
Wedge shaped contacts with opening angle $\Gamma=15^\circ$ are used in order to minimize the influence of the contacts on the superconducting properties
of the structures\cite{morelle02,morelle03}. The coherence length determined from a macroscopic co-evaporated sample was found to be $\xi(0)=156$~nm for
the disk and the three loops with a small opening. The thickness was $\tau=39$~nm. For the sample presented in Fig.~\ref{Fig:SEMLoops}(e), a coherence
length of $\xi(0)=120$~nm was determined in the same way as for the other structures. A thickness of $\tau=54$~nm was found from low angle X-ray
diffraction on a co-evaporated film and from AFM for the loop with $r_i=0.7~\mu m$. The $H-T$ phase diagram is constructed by four-point resistance
measurements using an ac transport current of 0.1~$\mu$A. The phase boundary is determined from a set of magnetoresistance curves measured at various
temperatures using a certain resistance criterion $R_c$.

\begin{figure*}
\centering
\includegraphics*[width=5.5cm,clip=]{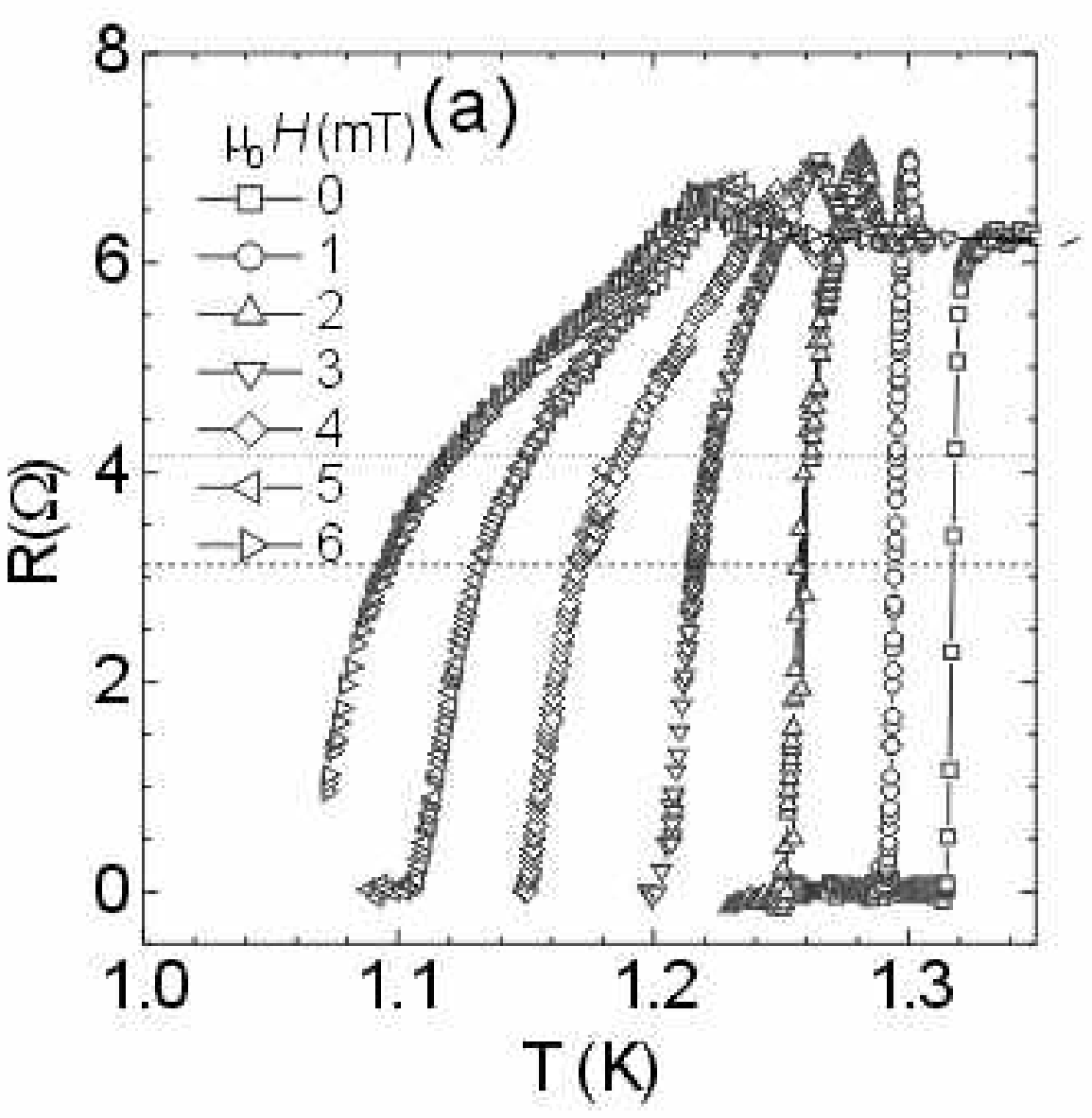}
\includegraphics*[width=5.5cm,clip=]{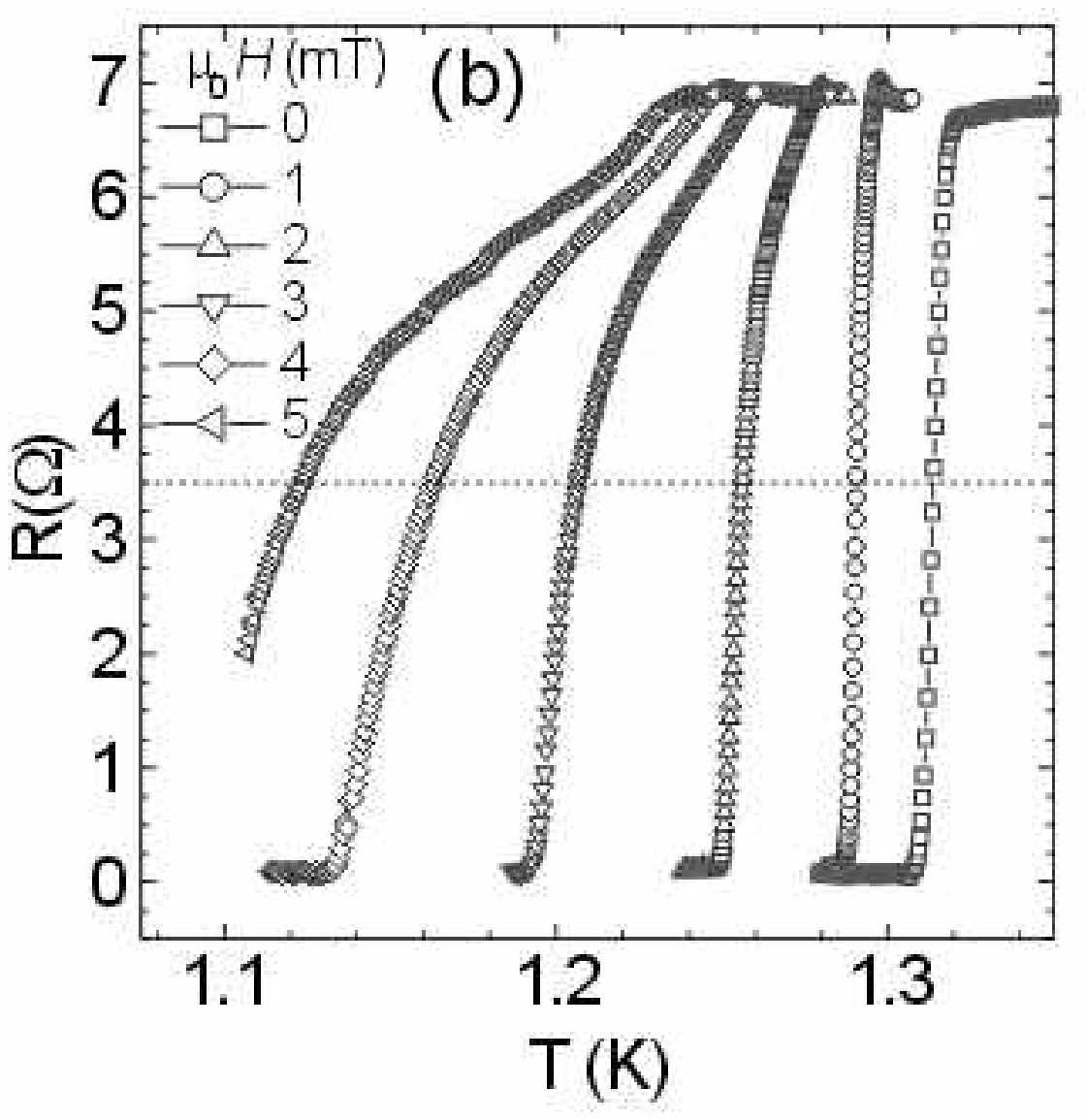}
\includegraphics*[width=5.5cm,clip=]{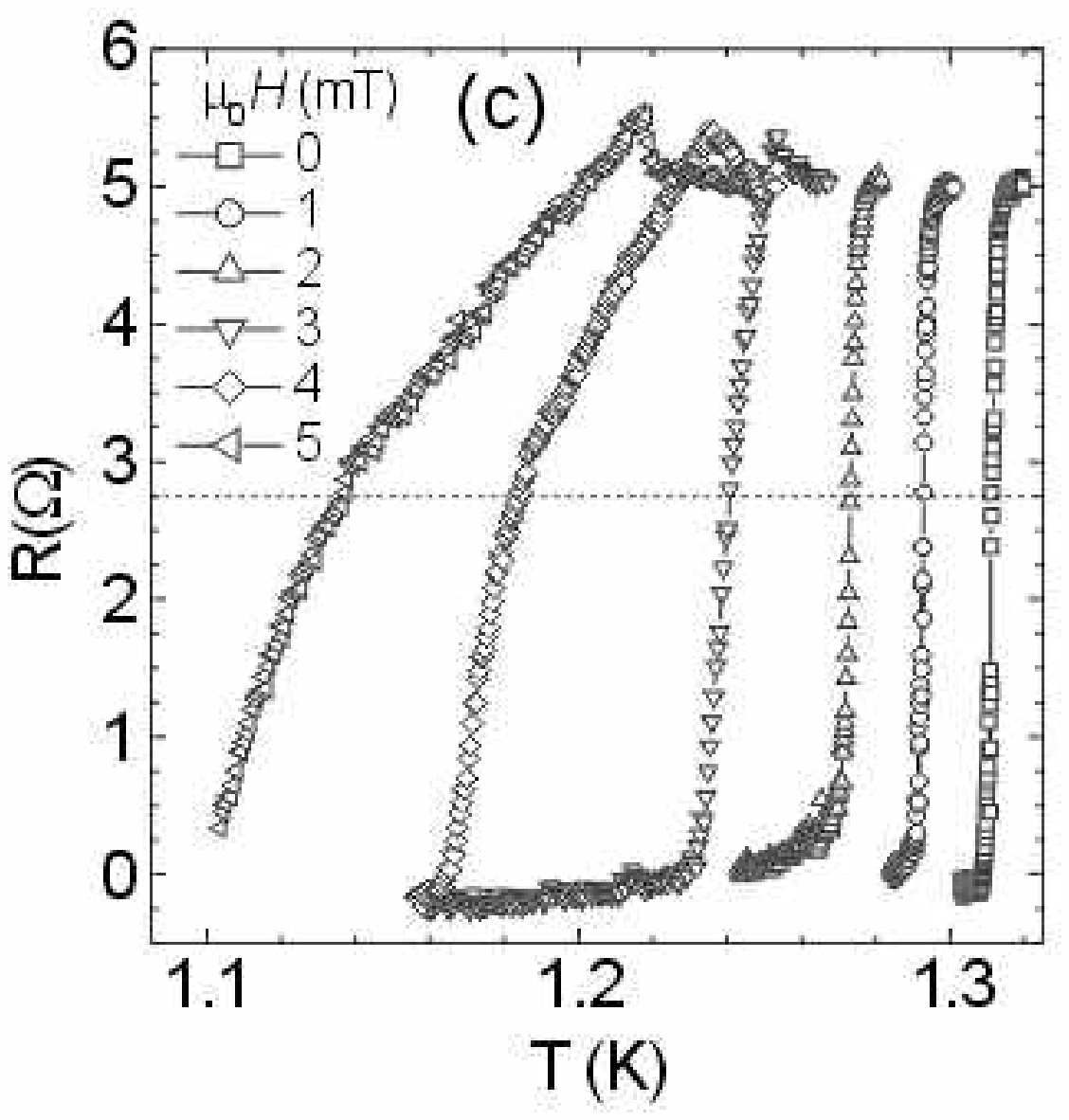}
\includegraphics*[width=5.5cm,clip=]{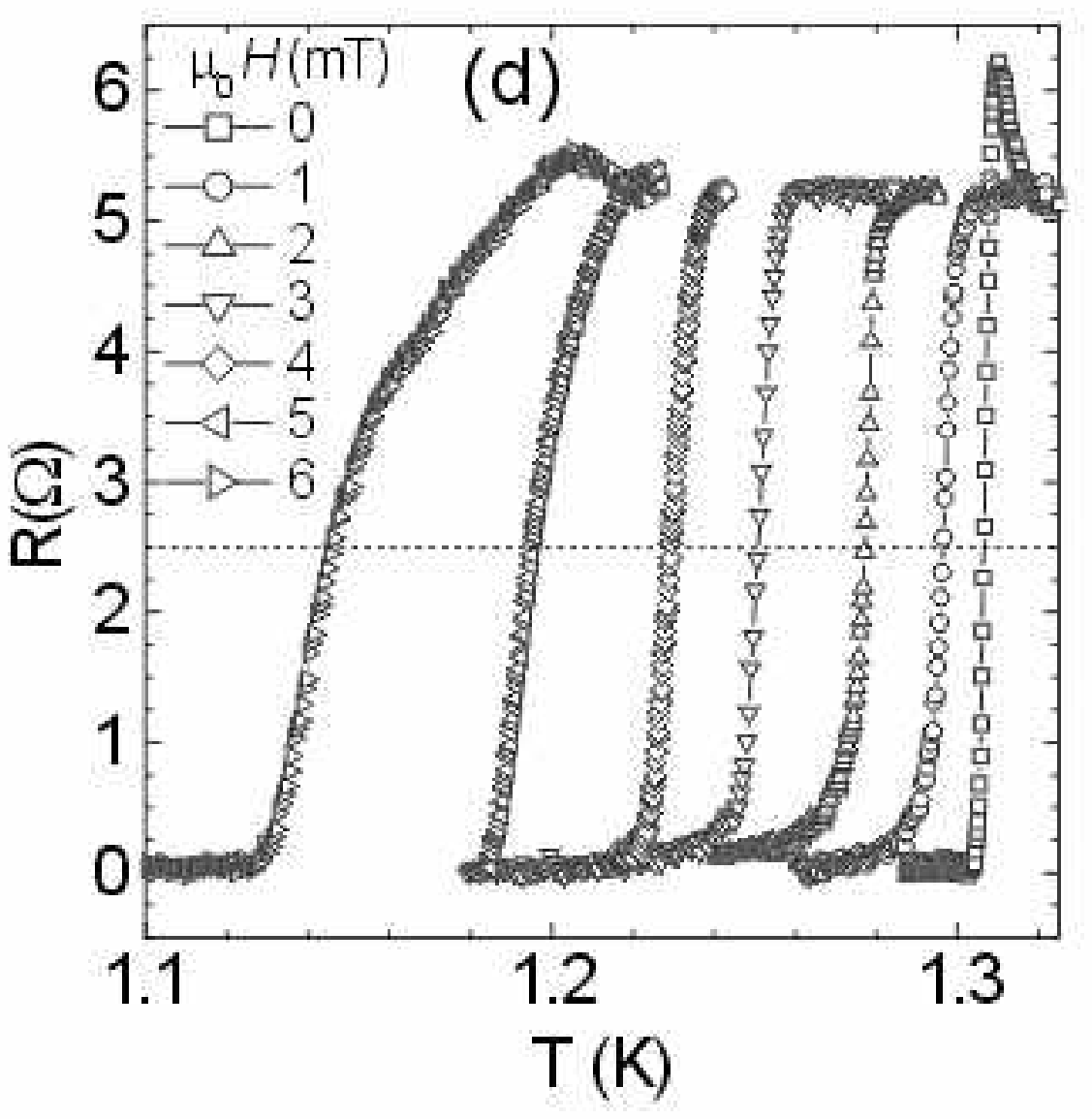}
\includegraphics*[width=5.5cm,clip=]{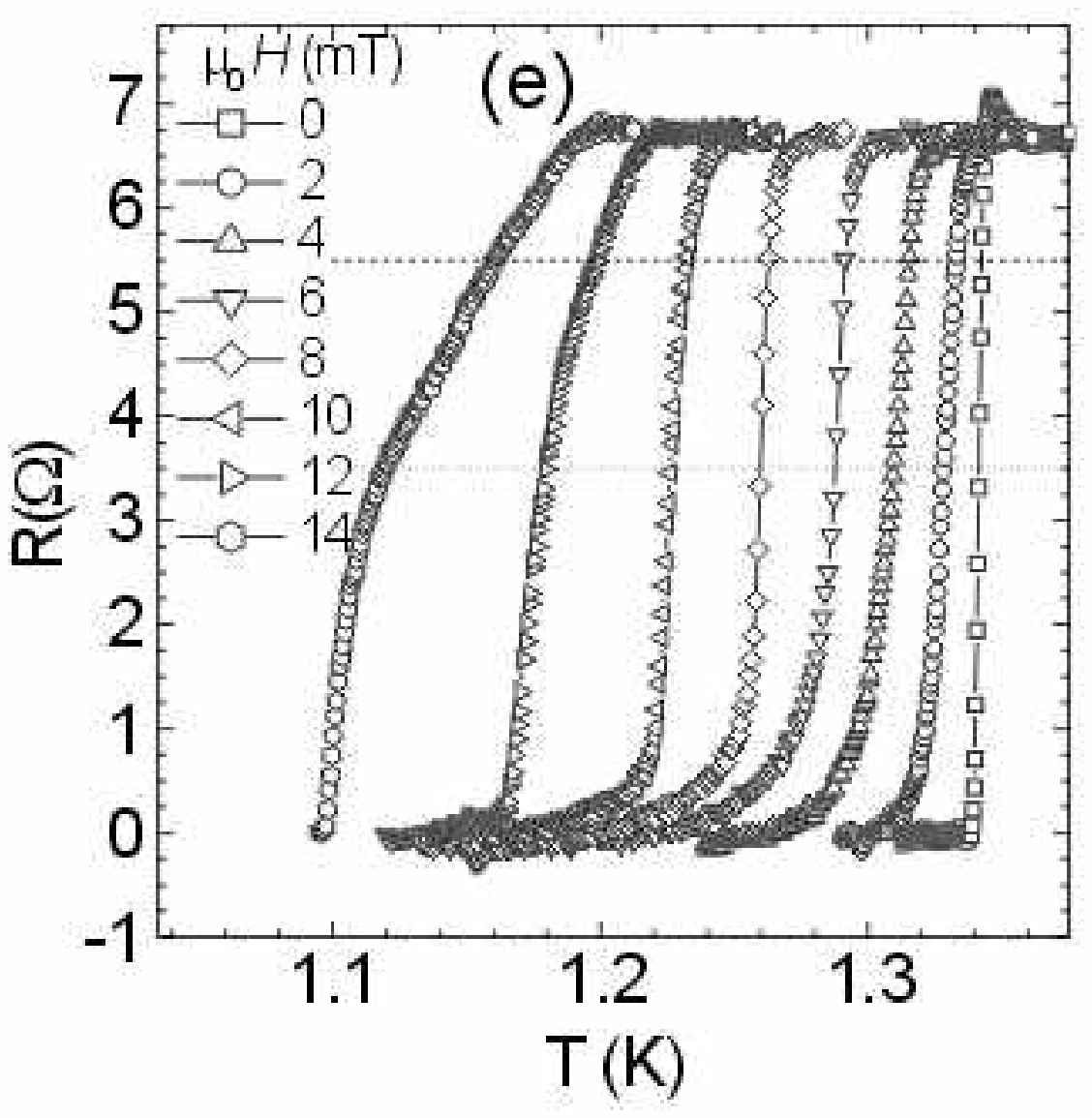}

\caption{Resistive transitions $R(T)$ for (a) a disk and for a loop with inner to outer radius ratio (b) $x=0.1$, (c) $x=0.3$, (d) $x=0.5$ and (e)
$x=0.7$ in different magnetic fields. The dashed and dotted lines show the resistance criteria used to determine the $T_c(H)$ phase boundary.}
\label{Fig:RTLoop}
\end{figure*}

\subsection{Resistance transitions}

The superconducting/normal resistance transitions for the disk and the rings with a different inner to outer radius ratio $x=r_i/r_o$ are shown in
Fig.~\ref{Fig:RTLoop}. The five different samples have a very similar temperature dependence of the resistance at different magnetic field as the samples
with wedge shaped contacts with opening angle $\Gamma=15^\circ$ presented in Ref.~\onlinecite{morelle03}. They are characterized by a slowly decreasing
resistance at high temperatures arising from the nucleation of superconductivity in the wedge contacts, followed by a sharp drop of the resistance once
superconductivity nucleates in the ring. Only small differences are seen in the amplitude of the resistance overshoot observed at certain magnetic
fields. It is probably due to small differences in the shape of the contacts that are responsible for the appearance or not of the resistance anomaly
created by a charge imbalance around superconducting/normal interfaces. The samples with $x$=0.3, $x$=0.5 and $x$=0.7 show a different behavior at low
magnetic fields. There, the situation is reversed. A sharp transition is firstly observed, followed by a broad transition at low resistance. We will show
below that the broad transition also corresponds to the nucleation in the wedges. This effect is observed in a broader magnetic field range when the
ratio $x$ increases.


A sharper transition at high magnetic fields is seen in the resistance transition if the thinnest ring. It is difficult to extract from the measurements
if this is caused by the smaller coherence length or by the geometry of the sample.

\subsection{$T_c(H)$ phase boundaries}

\begin{figure*}
\centering
\includegraphics*[width=5.5cm,clip=]{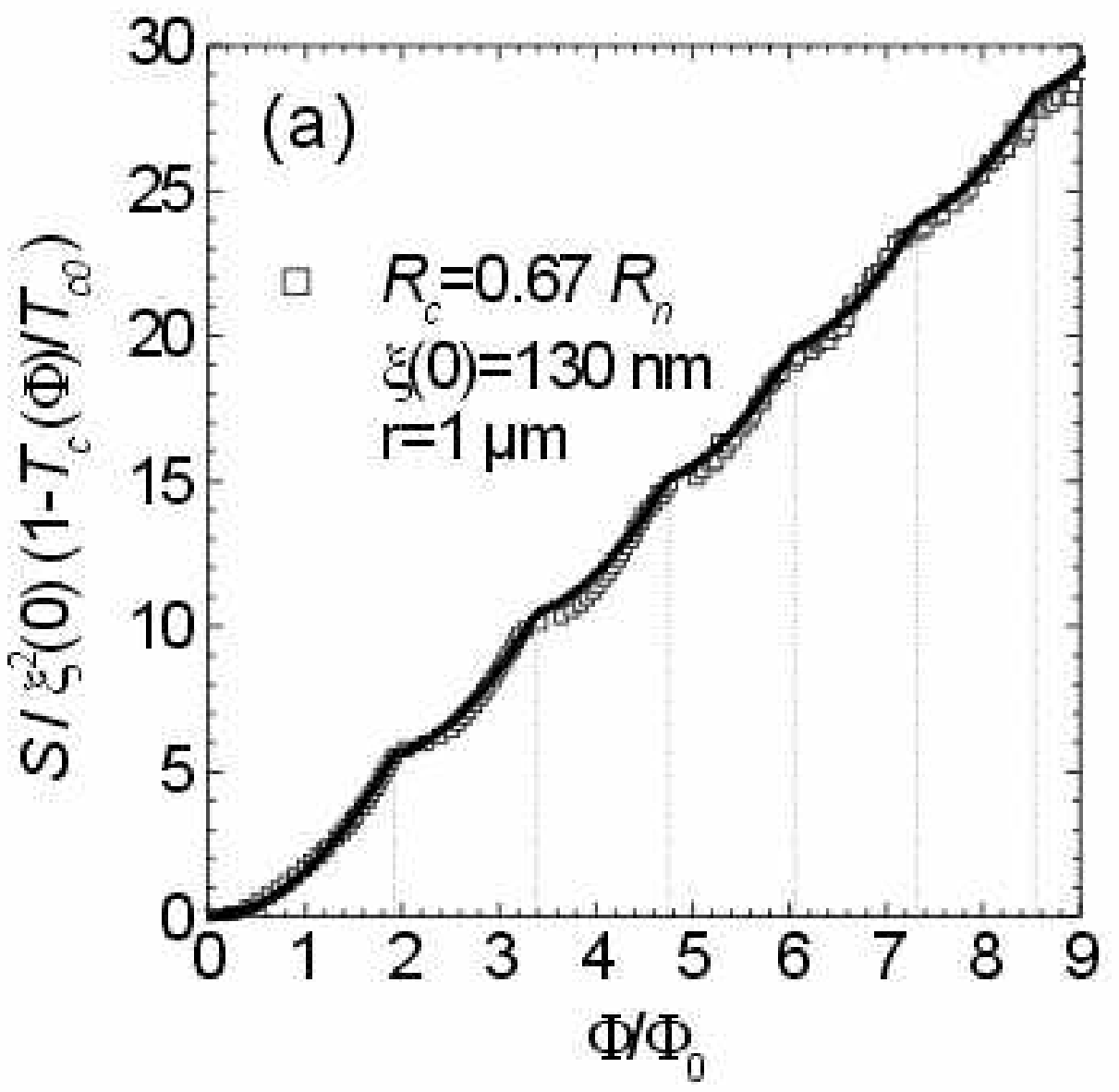}
\includegraphics*[width=5.5cm,clip=]{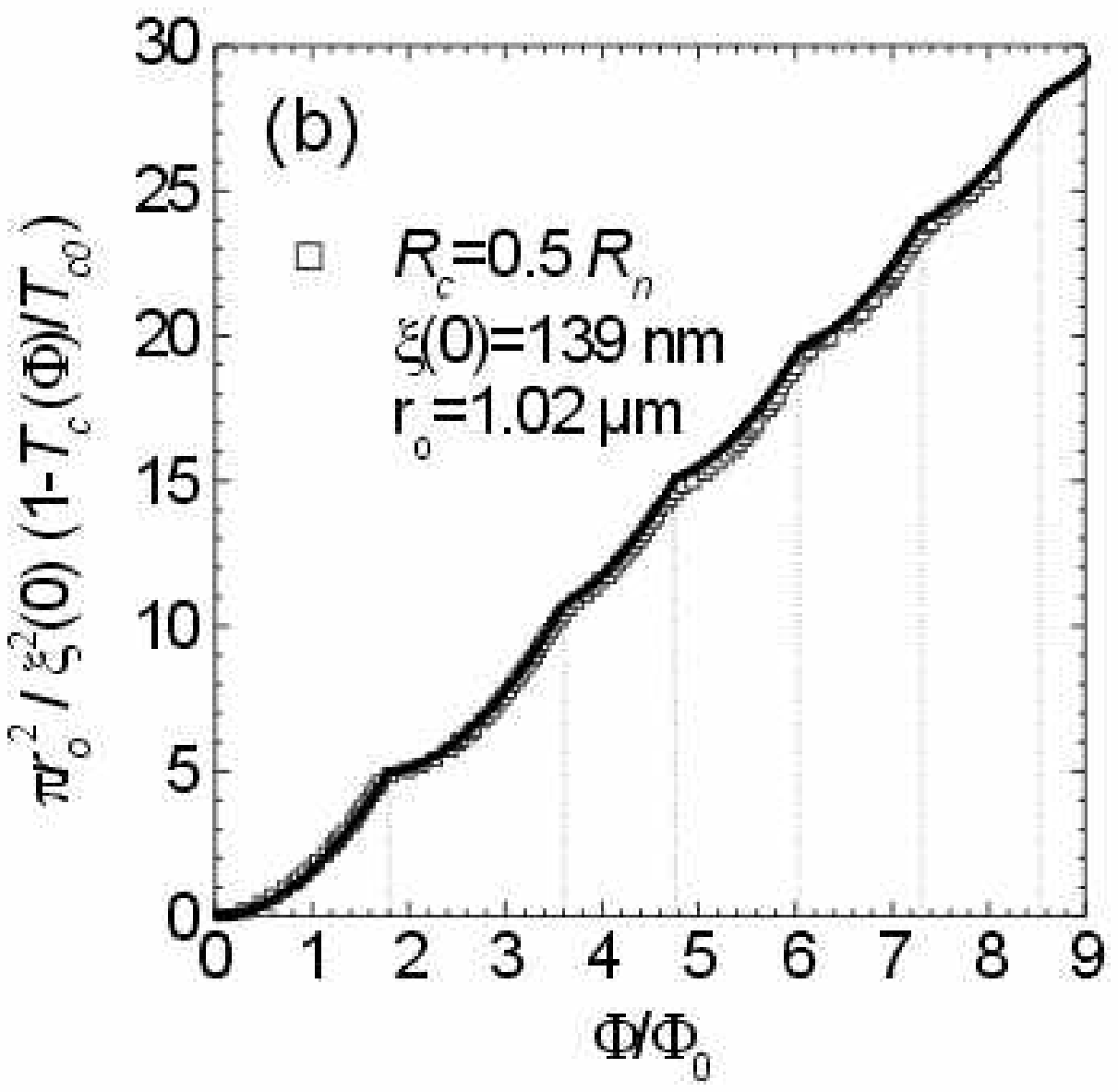}
\includegraphics*[width=5.5cm,clip=]{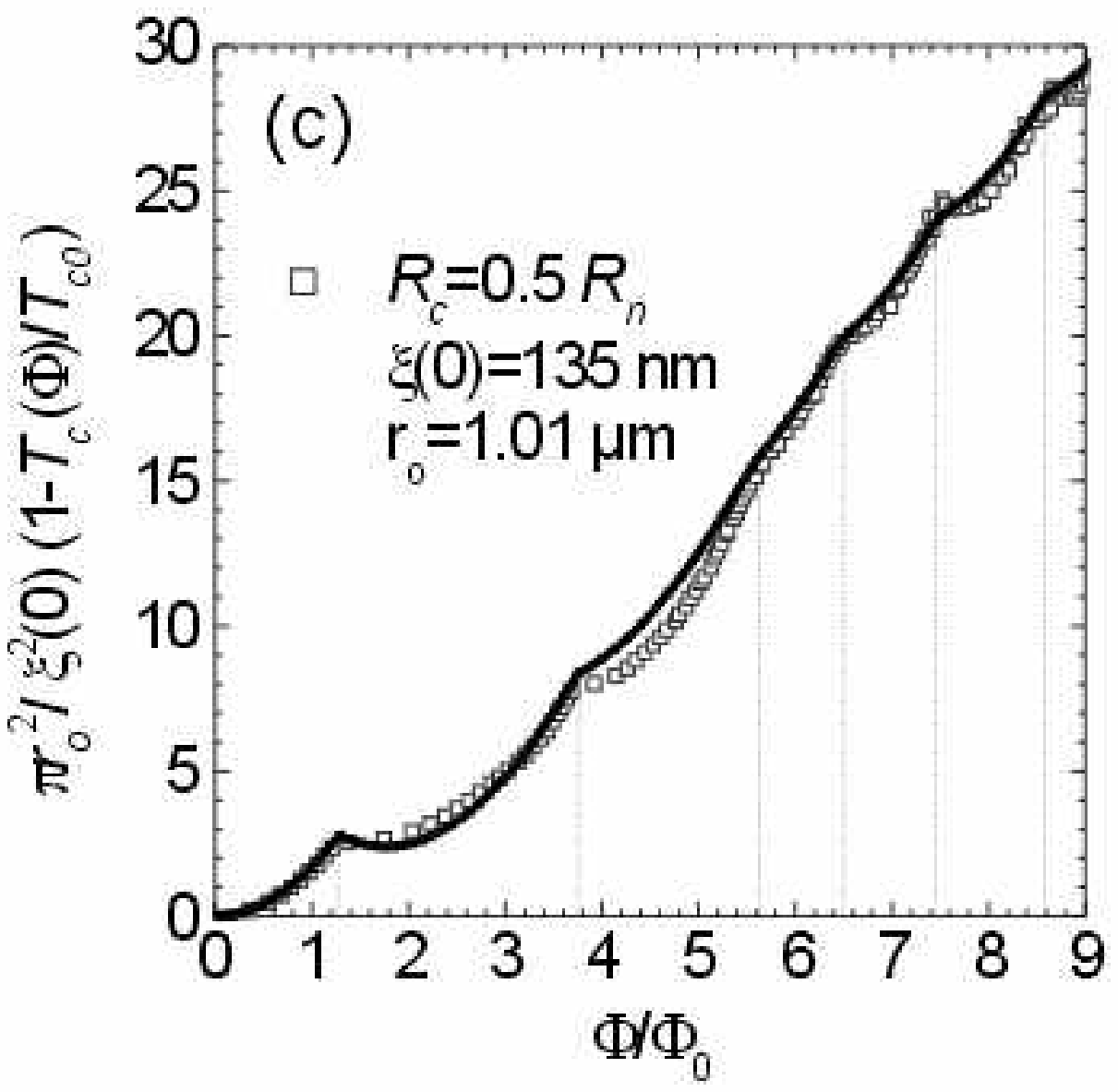}
\includegraphics*[width=5.5cm,clip=]{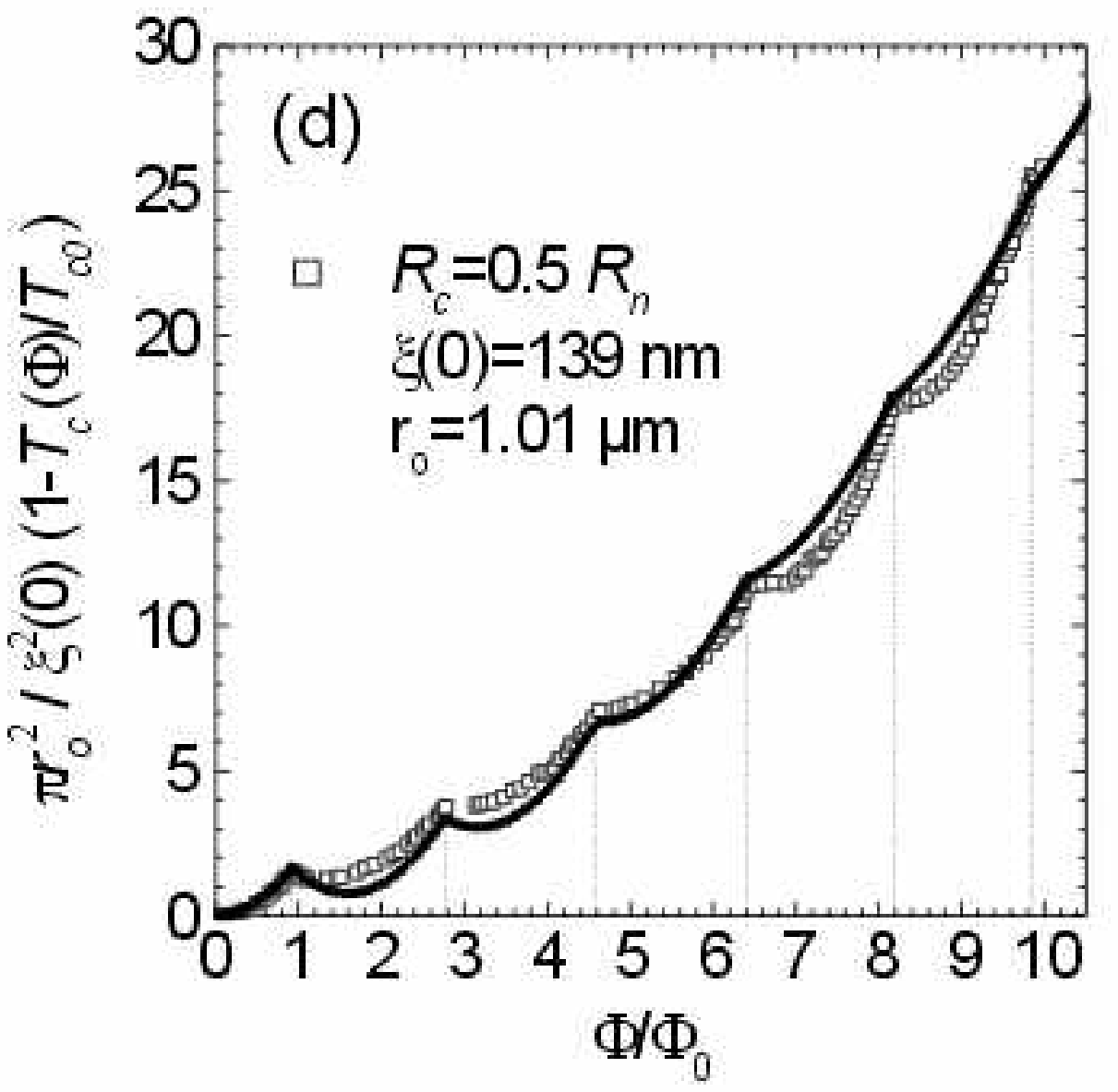}
\includegraphics*[width=5.5cm,clip=]{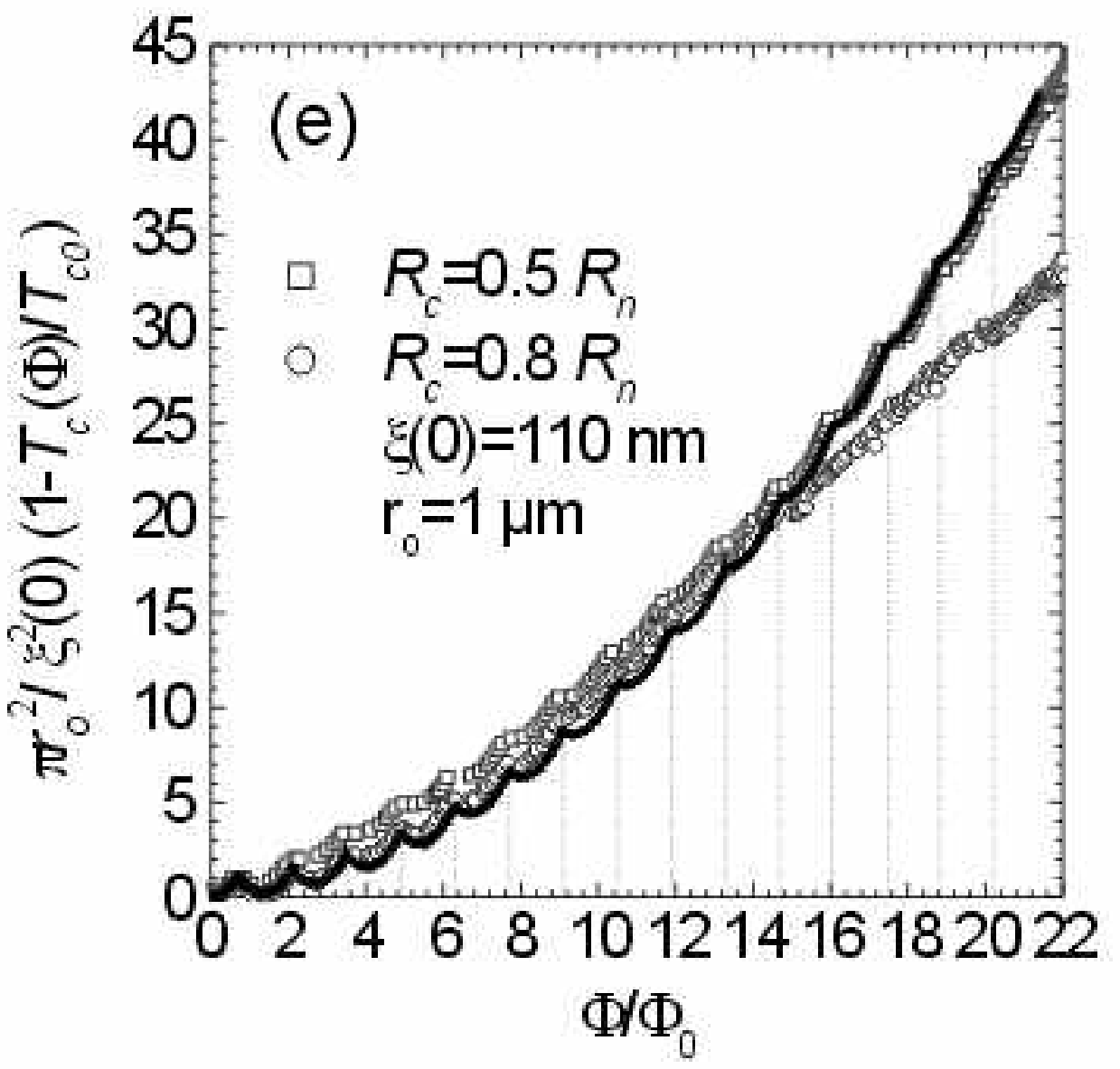}
\includegraphics*[width=5.5cm,clip=]{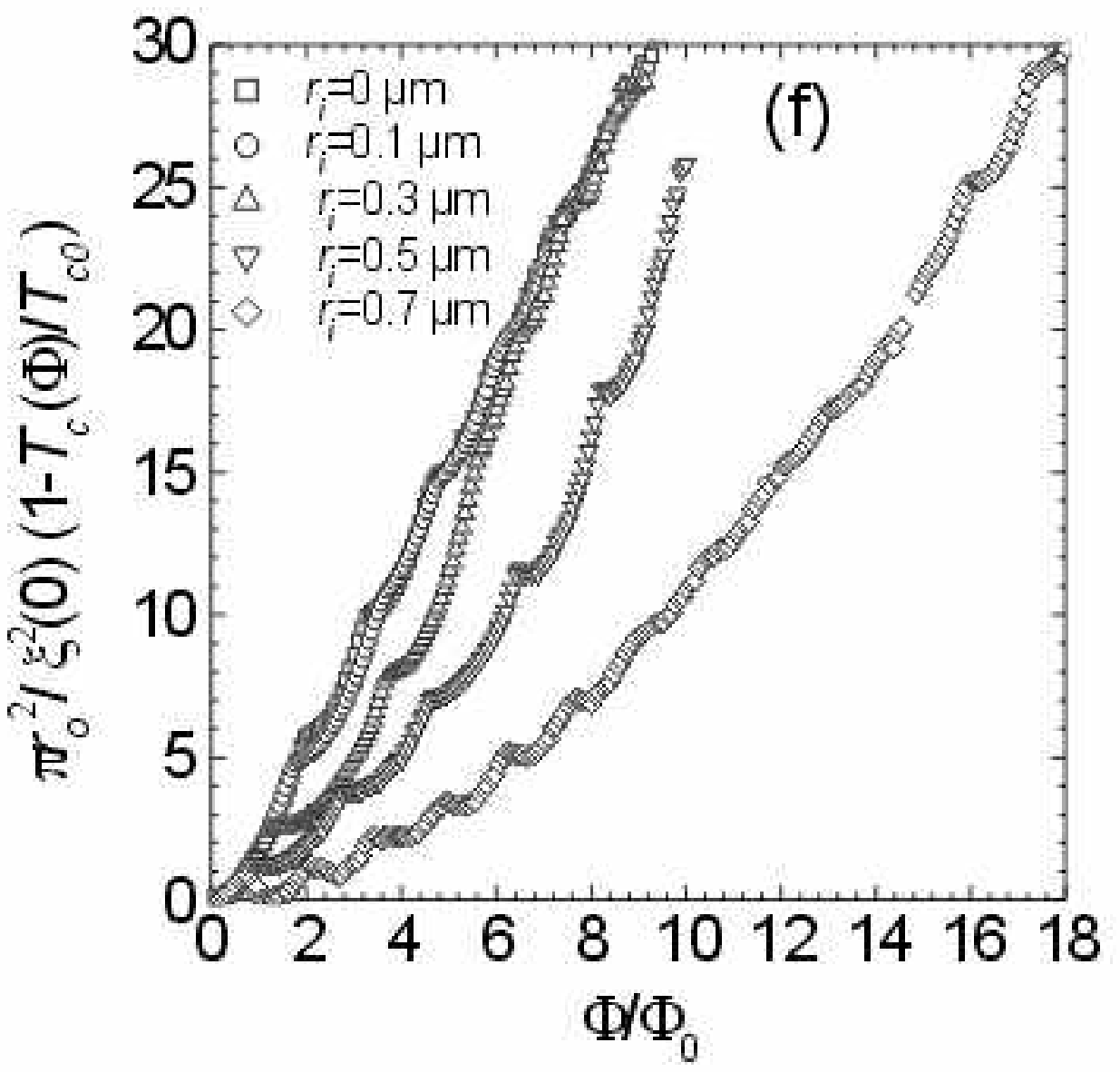}
\caption{Experimental $T_c(H)$ phase boundary of (a) a disk and for a loop with inner to outer radius ratio (b) $x = 0.1$, (c) $x = 0.3$, (d) $x = 0.5$
and (e) $x = 0.7$. The open squares [and the open circles in (e)] represent the measured data. The experimental $T_c(H)$ phase boundaries of the
different structures are compared in (f).} \label{Fig:TcBLoop}
\end{figure*}

The experimental phase boundary of the disk is presented in Fig.~\ref{Fig:TcBLoop}(a). The results are compared with theoretical calculations of the
nucleation field $H_{c3}^*(T)$ (full line in Fig.~\ref{Fig:TcBLoop}) by Bruyndoncx {\it et al.\/}\cite{bruyndoncx99crossover}.  A very good agreement
between the calculated and the measured curve is seen. Only a slightly lower coherence length ($\xi(0)=130$~nm) than the one found for a reference
macroscopic sample ($\xi(0)=156$~nm) had to be used for the experimental data.

The data for the ring with $x=0.1$ are shown in Fig.~\ref{Fig:TcBLoop}(b). The flux $\Phi$ on the field axis denotes the flux $\Phi=\mu_0 H \pi r_o^2$
through the ring and the hole. The $H-T$ diagram of the ring with the smallest hole resembles strongly the $T_c(H)$ line of the disk displayed in
Fig.~\ref{Fig:TcBLoop}(a). The phase boundary has a linear background superimposed with oscillations. A very good agreement between the measured and the
calculated curves is found.


Fig.~\ref{Fig:TcBLoop}(c) shows the $H-T$ diagram of the ring with $x=0.3$. Here, the linear dependence is only seen for vorticity $L>4$. At lower
magnetic field, a parabolic background suppression of $T_c$ is observed. The crossover from the linear to the parabolic regime occurs at $\pi
r_o^2/\xi^2(T) \approx 20$. This corresponds to a value $r_o-r_i \approx 1.8~\xi(T)$, which is in a good agreement with the thickness $\tau=1.84~\xi(T)$
for a crossover from a 1D to a 2D regime for a thin film in a parallel magnetic field\cite{schultens70zp, fink69crittick}.

A good agreement with the position of the cusps in the theoretical curve has been found. The amplitude of the oscillations in the experimental curve
deviates slightly from the calculated one. At $L$=1, between the first and the second $T_c(H)$ cusps, the experimental oscillation is less pronounced.
For higher vorticity, the opposite situation is seen where the amplitude of the experimental oscillations is larger than in the theoretical curve.

The penetration of the first vortex in the ring occurs at a lower magnetic field value than for the ring with the smallest hole [see
Fig.~\ref{Fig:TcBLoop}(b)], while the transitions $L=1\leftrightarrow 2$ to $L=5 \leftrightarrow 6$ occur at a higher magnetic field. That the
transitions take place at lower magnetic field value for a ring with thinner lines is expected since the transition between $L$ and $L+1$ occurs at
$\Phi/\Phi_0=L+1/2$ for an infinitely thin loop or cylinder. At higher magnetic fields, a giant vortex state is formed\cite{bruyndoncx99crossover} and
the disk with a small hole in the center behaves like the disk without hole. This, however, cannot fully explain why the change in vorticity is delayed
at high magnetic fields by introducing a small hole in a disk.

The measured $T_c(H)$ phase boundary of the ring with ratio $x=0.5$ is shown in Fig.~\ref{Fig:TcBLoop}(d). In the temperature range accessible with our
experimental setup, only a parabolic background dependence of the critical temperature on the magnetic field has been measured. By comparing the
experimental results with the calculations, a similar behavior as for the ring with $x=0.3$ is seen. The position of the cusps in the experimental curve
matches with the calculated transitions. However, no good agreement is found for the amplitude of the oscillations. For the vorticities $L=1$ and 2, the
amplitude is lower in the experimental curve, while for $L>3$, the amplitude is larger. At low $L$, the transition between states with different
vorticities occurs at a lower magnetic field than for the disk, while the transitions $L=3 \leftrightarrow 4$, $L=4 \leftrightarrow 5$ and $L=5
\leftrightarrow 6$ take place at a higher magnetic field, similar to what was observed for the ring with $x=0.3$.


The $H-T$ diagram of the ring with the thinnest line ($x=0.7$) is shown in Fig.~\ref{Fig:TcBLoop}(e). Two experimental curves are presented, one for
$R_c=0.5~R_n$ (open squares) and the second for $R_c=0.8~R_n$ (open circles). It can be seen that at a higher resistance criterion the parabolic
dependence switches to a linear regime at high magnetic field. For the curve calculated with the low resistance criterion, a quasi-parabolic background
suppression of $T_c(H)$ is observed over the whole measured range. The amplitude of the $T_c(H)$ oscillations is larger than in the samples with smaller
$x$ and the transition between states with different vorticities is almost periodic in field. A good agreement between the theoretical curve and the
experimental curve with $R_c=0.5~R_n$ is seen at high magnetic fields. At lower magnetic fields, a good agreement is found when using a higher resistance
criterion.

The phase boundaries of the four different loops are compared with the critical temperature of the disk in Fig.~\ref{Fig:TcBLoop}(f). All the curves
overlap with each other for $L=0$. \emph{It is interesting to note that an opening in the disk does not affect the phase boundary as long as no vortex is
trapped inside the superconductor.} Only the magnetic field range over which the state with $L=0$ exists at the phase boundary is lowered by introducing
a hole in the disk. The $T_c(H)$ line of the disk with the smallest hole in the middle does not deviate substantially from the phase boundary of the disk
without any opening. Only small changes in the position of the cusps is observed at low vorticity. For larger holes, the crossover from 2D to 1D regime
is clearly seen. The samples with the thinnest lines do not show the 2D regime in the studied temperature interval and only the parabolic dependence is
seen.

\begin{figure}
\centering
\includegraphics*[width=5.5cm,clip=]{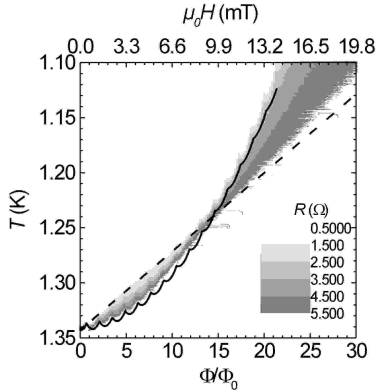}
\caption{Contour plot of the resistance $R(H,T)$ of a loop with $x=0.7$. The full line represents the calculated phase boundary of a loop with
$r_i=0.7~\mu m$ and $r_o=1~\mu m$, using a coherence length of 110~nm. The dashed line is the theoretical critical temperature of a wedge with opening
angle $\Gamma=15^\circ$ with $\xi=140$~nm.} \label{Fig:TBRLoop07}
\end{figure}

In order to reveal the origin of this different behavior at low and high magnetic field, a contour plot of the resistance $R(H,T)$ is presented in
Fig.~\ref{Fig:TBRLoop07}. Two different parts are clearly distinguished. Below 10~mT, the low resistance is linear, while the high resistance exhibits a
parabolic background superimposed with oscillations. Above 10~mT, the opposite situation occurs, where the low resistance has a parabolic decay with
small oscillations while the high resistance decreases monotonously. The parabolic part coincides with the nucleation of superconductivity in the loop
shown as a full line. The linear part arises from the nucleation in the wedge contacts.

By fitting the theoretical critical temperature of a wedge with opening angle $\Gamma=15^\circ$ to the linear part of the contour plot  (dashed line), a
coherence length $\xi(0)=140$~nm is obtained. This differs from the coherence length $\xi(0)=110$~nm that was used to find a good agreement between the
experiment and the theoretical curve of a loop. A possible origin of this discrepancy could be a width of the loop that has been evaluated to be smaller
than the real size. An estimate of the thickness that would satisfy the coherence length used for the calculation of the wedge contacts can be obtained
from the analysis of the nucleation field of a thin wire of a film in a parallel magnetic field. From the calculation of the nucleation field of a thin
film in a parallel field\cite{tinkbook}, a value for the width of the loop of $0.38~\mu m$ is obtained instead of $0.3~\mu m$ found from SEM
measurements. This difference is too large to be explained only by an error in the characterization of the sample. The opening angle of the contacts can
be determined with a high accuracy so that a divergence arising from a wrong determination of $\Gamma$ could be excluded. It means that either the
nucleation of superconductivity is delayed in the wedges due to the presence of the loop or that the nucleation in the loop is enhanced by the contacts.
It is also possible that the coherence length in the loop is slightly different from that in the wedge. The sample geometry can indeed affect the
superconducting parameters $\lambda$ and $\xi$ in a structure of mesoscopic size similar to the case of a thin film where the effective penetration depth
increases as $\lambda '= \lambda^2/\tau$, taking into account the demagnetization effects. The renormalization of $\lambda$ and $\xi$ should therefore be
calculated in a self-consistent way from the sample geometry.

The shape of the resistive curves in Fig.~\ref{Fig:RTLoop}(e) can be easily understood from Fig.~\ref{Fig:TBRLoop07}. It was clearly seen that in low
magnetic fields the nucleation first occurs in the ring and is then followed by the nucleation in the contacts. Due to the different field dependence of
the $T_c(H)$ of the ring and the contacts, the opposite occurs in higher magnetic fields. Two different shapes are therefore distinguished in the
resistive curves depending on the part where superconductivity starts to nucleate. The same happens in the rings with $x=0.3$ and $x=0.5$ since $T_c$
also has a parabolic field dependence for low fields so that the broad transition at low resistance seen in Figs.~\ref{Fig:RTLoop}(c)
and~\ref{Fig:RTLoop}(d) is due to the nucleation of superconductivity in the wedge that takes place after the nucleation in the ring at low magnetic
fields. The normal parts of the sample can however partially become superconducting by the proximity effect with the neighboring superconducting part.

\subsection{Resistance criterion}

\begin{figure*}
\centering
\includegraphics*[width=5.5cm,clip=]{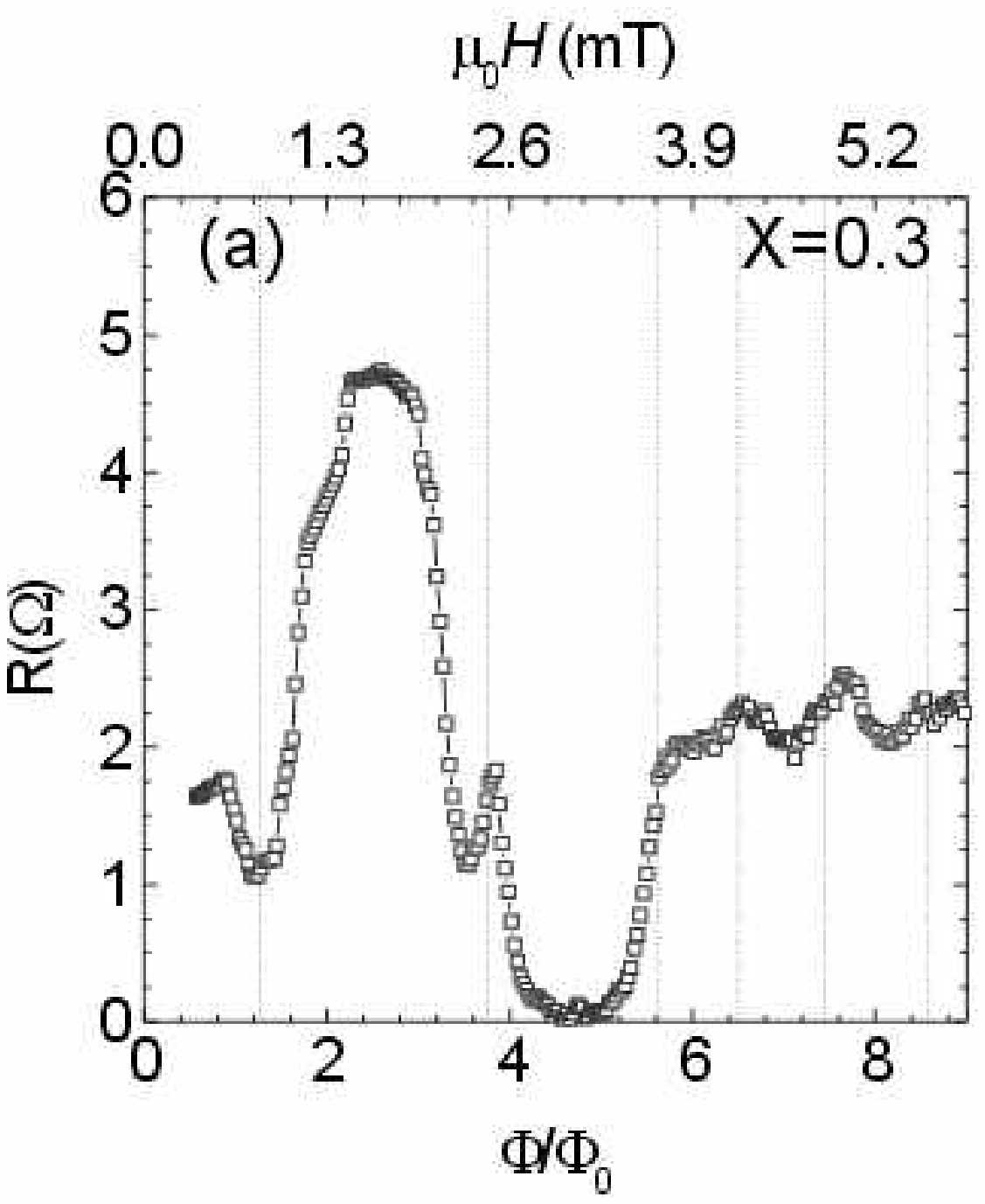}
\includegraphics*[width=5.5cm,clip=]{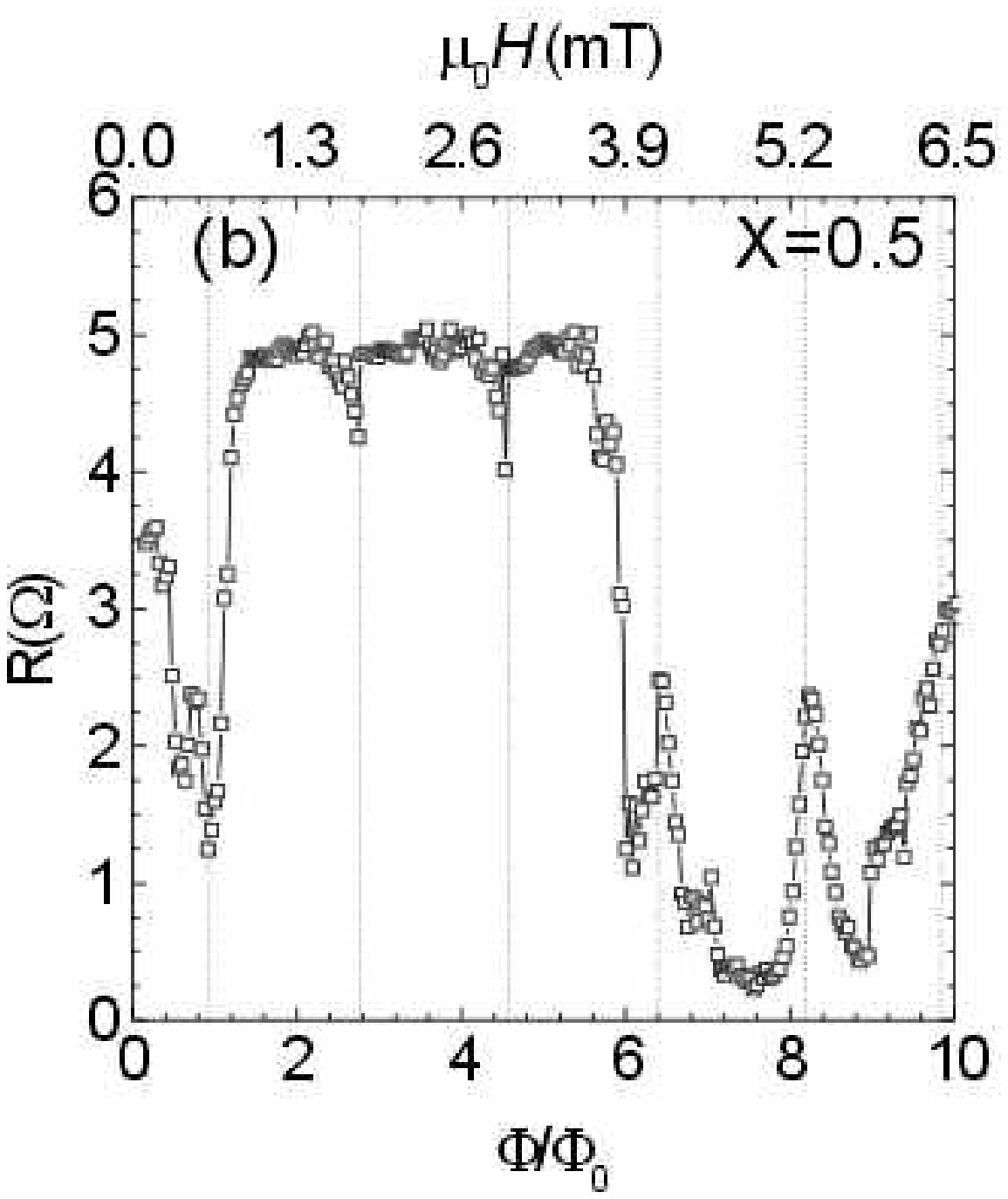}
\includegraphics*[width=5.5cm,clip=]{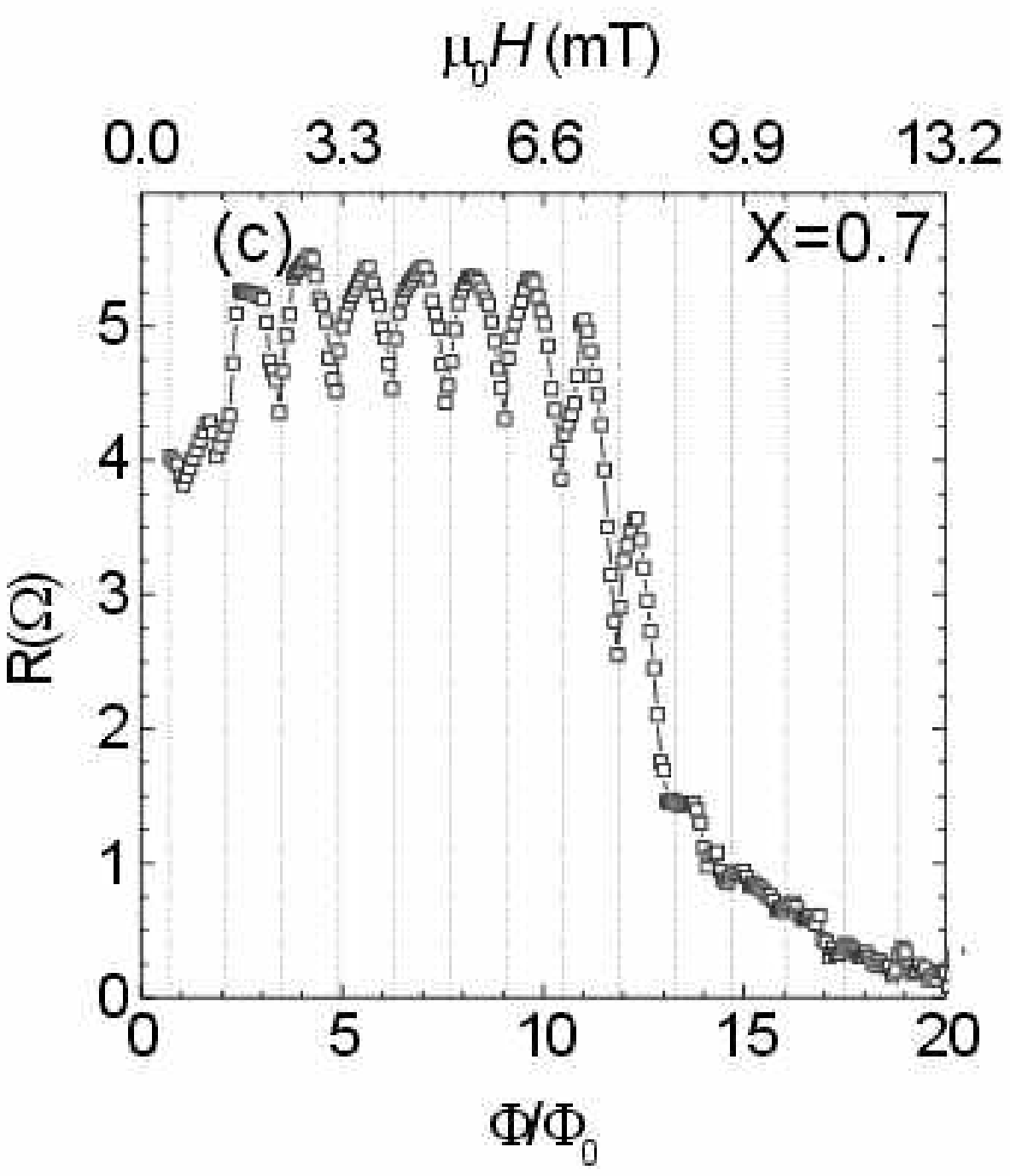}
\caption{Resistance $R(H,T)$ of a loop with (a) $x=0.3$, (b) $x=0.5$ and (c) $x=0.7$ measured at a temperature $T(H)=T_c(H)$ following the theoretical
phase boundary $T_c(H)$ shown in Fig.~\ref{Fig:TcBLoop}(c), (d), and (e), respectively. The dotted lines represent the position of the cusps in the
theoretical $H-T$ diagram.} \label{Fig:RcritLoop}
\end{figure*}

Since the contacts have a different field dependence than the studied structures, the choice of a constant resistance criterion for the determination of
the $T_c(H)$ phase boundary seems not to be obvious anymore. In order to study the best resistance criterion that should be used, the experimental
resistance $R(H,T)$ was determined at the temperature $T(H)$ corresponding to the theoretically calculated phase boundary $T_c(H)$. The results are given
in Fig.~\ref{Fig:RcritLoop} for the loops with $x=0.3$, $x=0.5$ and $x=0.7$. The curve corresponding to the resistance of the ring with the thinnest hole
($x=0.1$) is not shown. These curves show the expected behavior with an approximately constant resistance equal to the resistance criterion used for the
determination of $T_c(H)$ presented in Fig.~\ref{Fig:TcBLoop}(b). For other rings a similar $R(H,T)$ dependence was obtained. At low magnetic field, the
resistance is high, and it drops to a low value above a certain magnetic field. The field where this transition occurs increases with increasing $x$ and
corresponds approximately to the position where the phase boundary of the contacts crosses the phase boundary of the loop. For the ratio $x=0.3$, a
normal regime is found at high magnetic field with an almost constant resistance with a value approximately equal to the resistance criterion used for
the determination of the phase boundary. It occurs when the linear regime is recovered. This is not seen in the two other samples since there, the linear
regime is not attained. At low magnetic field, superconductivity nucleates first in the ring. This is the upper part in the resistance curves. At this
point, a high resistance criterion should be taken. For higher magnetic fields, the resistance starts to drop once the contacts become superconducting.
In order to determine the $H-T$ diagram of the ring in that region, a low resistance transition should be taken. This is exactly what is seen in
Fig.~\ref{Fig:RcritLoop}. The exact shape of the curves shown in the figures strongly depends on the coherence length used for the calculation of the
theoretical $T_c(H)$ line, but the general behavior will not strongly change while using a slightly different value of $\xi(T)$.

The origin of the discrepancy in the amplitude of oscillations in the $T_c(H)$ phase boundary is most probably due to the fact that \emph{a constant
criterion works well to determine the phase boundary when the critical temperature of the contacts has a similar field dependence than the studied sample
geometry.} When this is not the case, the determination of the phase boundary is strongly hindered.

It is also interesting to note that oscillations are present in the curves, with a maximum at the transition between different vorticities which becomes
a minimum at high magnetic fields. This crossover also corresponds approximately to the field where the $T_c(H)$  curves of the loops and of the contacts
cross each other. A minimum in the resistance curves of Fig.~\ref{Fig:RcritLoop} is observed when the difference in critical temperature of the loop and
of the wedge contacts is minimal and a maximum when the difference is maximal. This also reflects the observed difference in amplitude of the
oscillations in the experimental and theoretical phase boundaries.

Meyers~\cite{meyers03} calculated that the order parameter evolves from stronger at the inner part to stronger at the outer part when the vorticity is
increased. This could affect our measurements since our applied transport current would flow along the inner edge at fields slightly lower than the field
where the transition of vorticity happens and then flow along the outer edge for slightly higher fields. Since our contacts are situated on the outer
edge, a higher resistance could be expected when the transport current is flowing along the inner edge. We however do not see an increase (decrease) of
the resistance just before (after) the cusp.

\section{Non-symmetric geometries}

In this section the nucleation of superconductivity is studied for disks with a hole. The aim of this study is to analyze the effect of displacing the
hole from the middle of the disk on the phase boundary.

\label{sec:nonsym}

\begin{figure}
\centering
\includegraphics*[width=8.5cm,clip=]{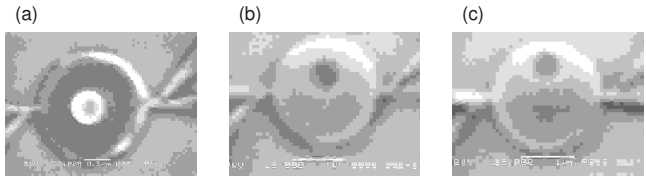}
\caption{SEM micrograph of an Al loop with outer radius $r_o$=1~$\mu m$ and inner radius $r_i$=0.3~$\mu m$ with the hole (a) in the center, (b) moved
over a distance $a=0.3~\mu m$ and (c) over $a=0.6~\mu m$ from the center.} \label{Fig:SEMDecenters}
\end{figure}

A SEM micrograph of the studied structures is presented in Fig.~\ref{Fig:SEMDecenters}. The three samples are the disks with outer radius $r_o = 1~\mu m$
and contain a circular opening with radius $r_i= 0.3~\mu m$. The hole is in the center of the disk [Fig.~\ref{Fig:SEMDecenters}(a)] or is displaced from
the center of the disk over a distance $a=0.3~\mu m$ [Fig.~\ref{Fig:SEMDecenters}(b)] and $a=0.6~\mu m$ [Fig.~\ref{Fig:SEMDecenters}(c)]. The two
asymmetric samples were co-evaporated in the same run as the circular symmetric ring with ratio $x=0.3$ discussed in Section~\ref{sec:rings}. They have
the same thickness $\tau=39$~nm. The coherence length of $\xi(0)= 156$~nm was determined from a reference macroscopic film. Wedge shaped current and
voltage contacts with an opening angle $\Gamma=15^\circ$ were used.



\begin{figure*}
\centering
\includegraphics*[width=5.5cm,clip=]{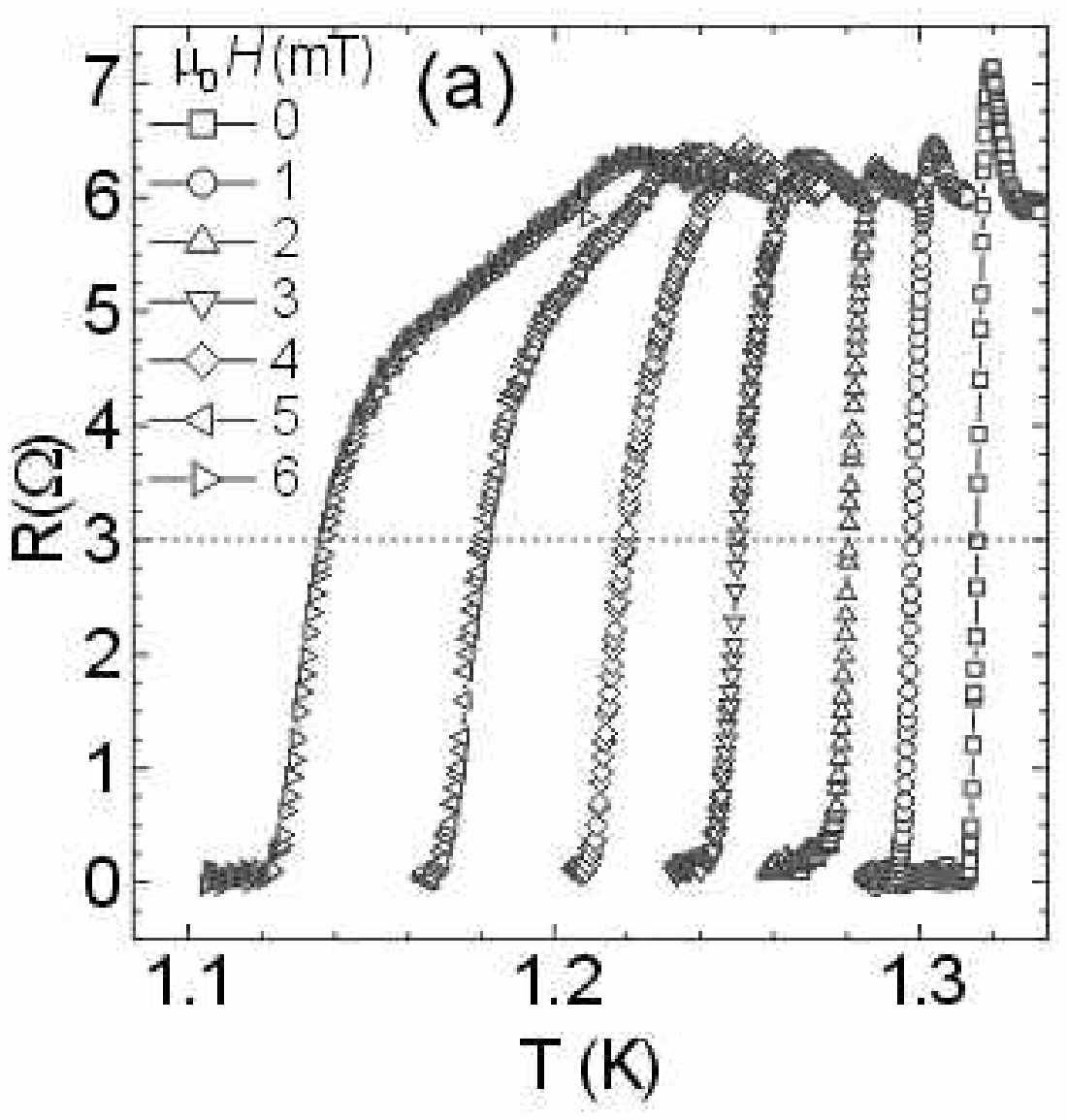}
\includegraphics*[width=5.5cm,clip=]{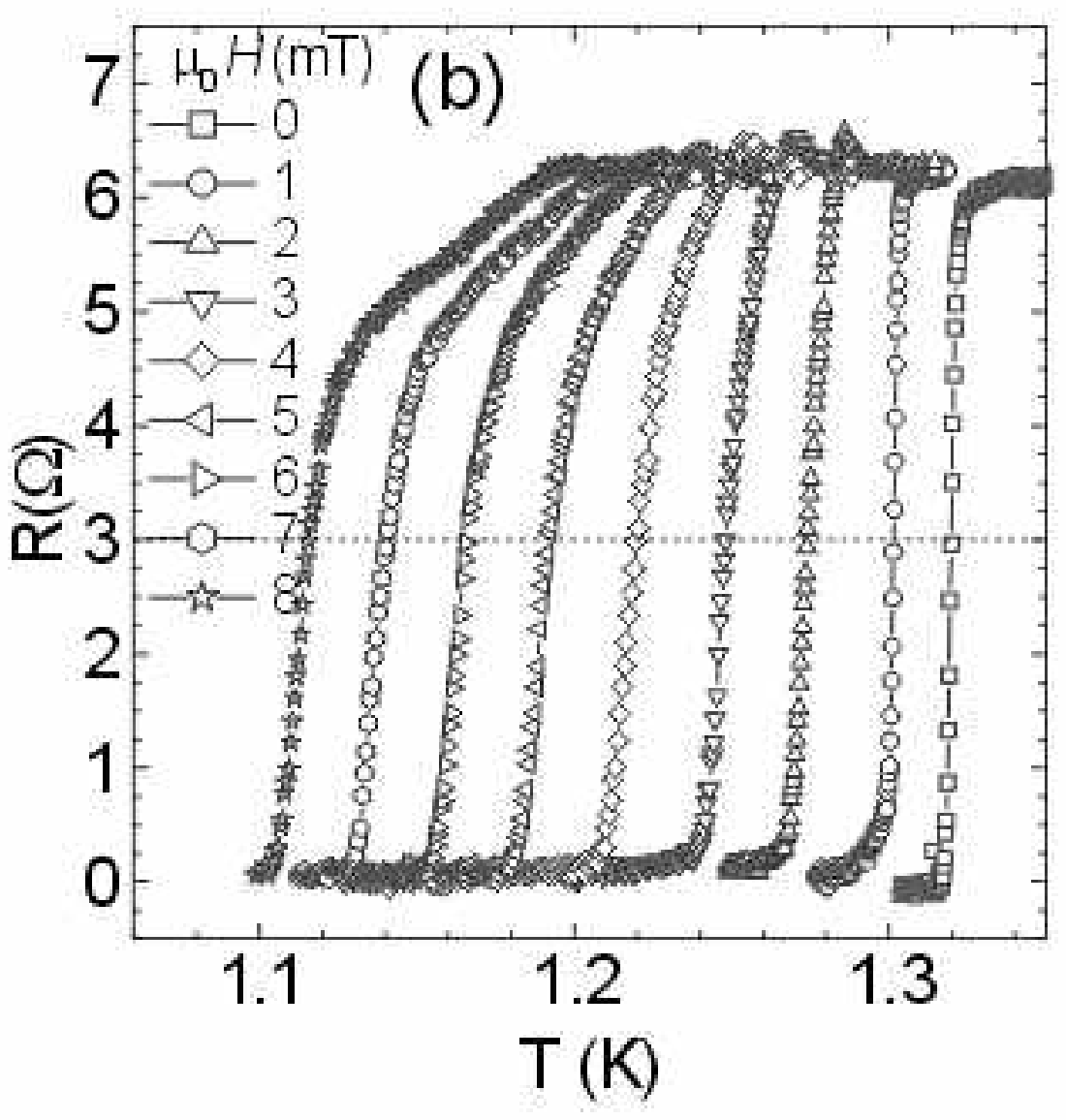}
\caption{Resistive transitions $R(T)$ in different magnetic fields for a ring with outer radius $r_o$=1~$\mu m$ and with a hole radius $r_i$=0.3~$\mu m$.
The hole is  moved (a) by 0.3~$\mu m$ and (b) by 0.6~$\mu m$ from the center. The dashed line shows the resistance criterion used to determine the
$T_c(H)$ phase boundary.} \label{Fig:RTDecentr}
\end{figure*}

The resistive transitions of the rings with $a=0.3~\mu m$ and $a=0.6~\mu m$ are shown in Fig.~\ref{Fig:RTDecentr}. The transitions at 2~mT for $a=0.3~\mu
m$ and at 1 and 2~mT for $a=0.6~\mu m$ exhibit a sharp drop by decreasing the temperature followed by a slowly decaying resistance in the lowest part of
the curves. This is similar to the curves of the symmetric ring shown in Fig.~\ref{Fig:RTLoop}(c), but less pronounced. It was seen in
Section~\ref{sec:rings} that the sharp decrease of $R$ is due to the nucleation of superconductivity that occurs first in the ring at these  magnetic
fields. For higher fields, the curves show transitions that are similar to the $R(T)$ curves of the disk. The sharp part at $\mu_0 H \gtrsim 3$~mT also
corresponds to the nucleation of superconductivity in the ring. At higher magnetic fields, the nucleation starts first in the wedges and is then followed
by the ring.

\begin{figure*}
\centering
\includegraphics*[width=5.5cm,clip=]{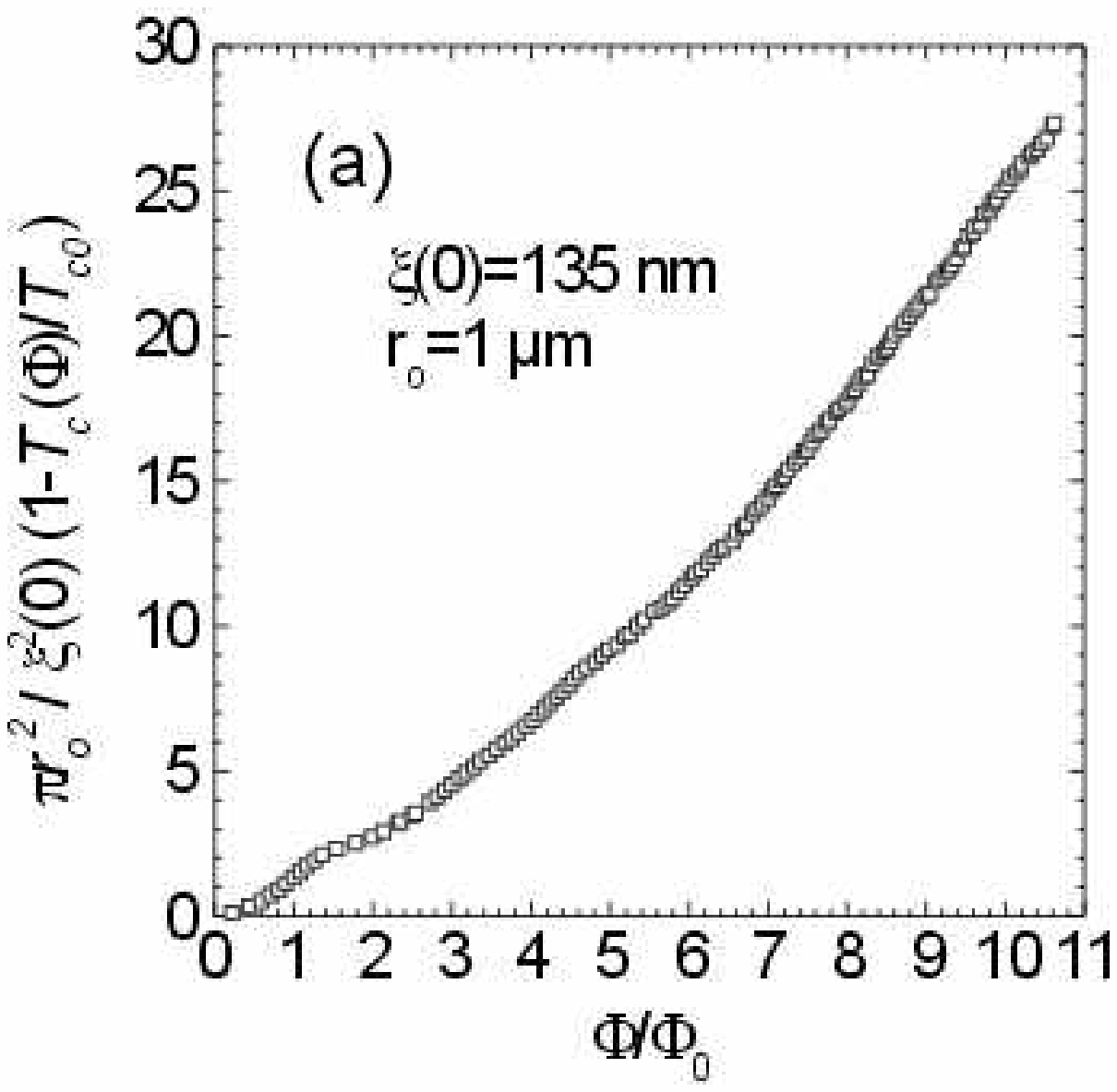}
\includegraphics*[width=5.5cm,clip=]{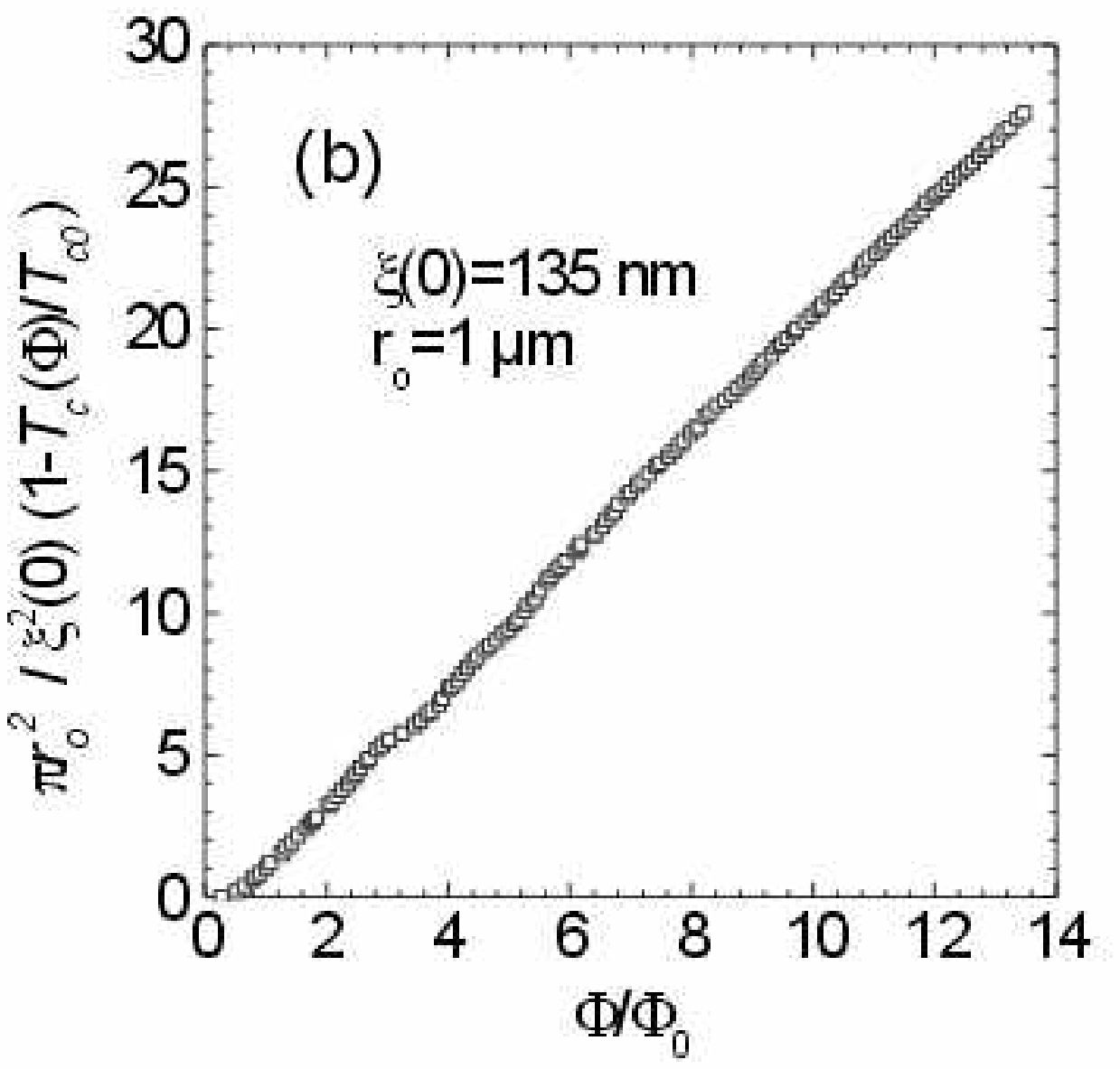}
\includegraphics*[width=5.5cm,clip=]{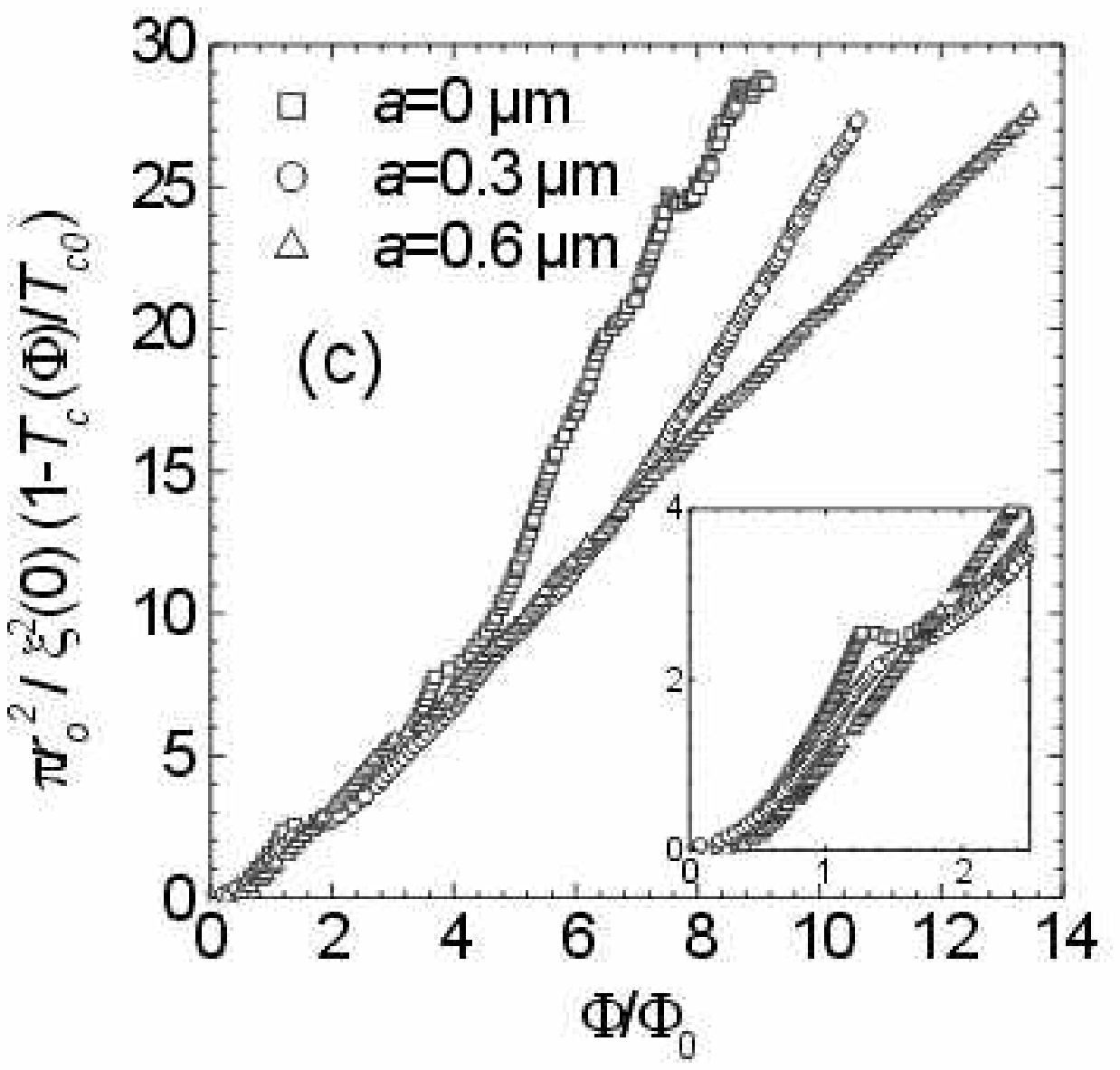}
\caption{Experimental $T_c(H)$ phase boundary of a loop with outer radius $r_o$=1~$\mu m$ and with a hole radius $r_i$=0.3~$\mu m$ determined for a
resistance criterion of 1/2~$R_n$. The hole is moved (a) by 0.3~$\mu m$ and (b) by 0.6~$\mu m$ from the center. The open squares represent the data
normalized by the coherence length $\xi(0)$=135~nm. (c) Comparison between the different phase boundaries. The hole is in the center (open squares) or
moved by 0.3~$\mu m$ (open circles) and 0.6~$\mu m$ (open triangles) from the center. The inset shows a magnification of the low magnetic field region.}
\label{Fig:TcBDecentr}
\end{figure*}

The $H-T$ diagram of the disk with an off-centered hole displaced by $a=0.3~\mu m$ off the center is shown in Fig.~\ref{Fig:TcBDecentr}(a). Small
oscillations are seen in the phase line. A behavior in between the parabolic and the linear field dependence is observed. At higher magnetic field, the
oscillations are almost not distinguishable anymore. The phase boundary of the disk with the hole displaced by $a=0.6~\mu m$ from the center is given in
Fig.~\ref{Fig:TcBDecentr}(b). In the temperature range accessible in our experimental setup, only a linear regime was observed. Also very weak
oscillations were distinguished.

The two  $T_c(H)$ lines are compared with the phase boundary of the circular symmetric ring, with the same inner and outer radii, shown in
Fig.~\ref{Fig:TcBDecentr}(c). The three curves have approximately the same behavior at low magnetic field, except that the oscillations are less
pronounced for the non-symmetric rings. It was seen in Section~\ref{sec:rings} that an opening does not affect the phase boundary for $L=0$. There, the
symmetry was kept. When the circular symmetry is broken, the phase boundary is strongly affected as can be seen in the inset of
Fig.~\ref{Fig:TcBDecentr}(c) even when no vortices are trapped in the sample. At high magnetic fields, the curves separate. Increasing the asymmetry
enhances superconductivity. Baelus~{\it et al.\/}\cite{baelus00} calculated the free energy, the magnetization and the Cooper-pair density of
non-symmetric rings with finite width. They found that the density of the superconducting condensate was the highest in the narrowest region of the
superconductor when $L \neq 0$. They argued that the trapped flux tries to restore the broken symmetry. That superconductivity is stronger in the
smallest part of the sample is probably due to the fact that the critical field is enhanced in thin lines. With the configuration of the contacts that
was used (see Fig.~\ref{Fig:SEMDecenters}), a superconducting `bridge' can be formed across which the external current applied for transport measurements
can pass. The critical field of this area will probably be higher than in the one with the largest area of superconducting material. The measured phase
boundary is therefore most likely only the phase boundary of the bridge and not of the full sample. Given that no supercurrent can circulate around the
opening, \emph{a singly connected state is then recovered}\cite{berger95prl,berger97prb}. Since a superconducting path will always be found across the
bridge, a lower resistance criterion for $T_c(H)$ will not determine the nucleation in the whole sample. The phase boundary of the complete structure
could only be probed with contacts turned by 90$^\circ$.

It is worth emphasizing that the phase boundary of the structure with $a=0.6~\mu m$ exhibits a linear field dependence, typical for a 2D behavior. The
thinnest part of the sample where the nucleation firstly occurs can be seen as a curved line of varying width. When the asymmetry is less pronounced
($a=0.3~\mu m$), a curve in between the parabolic and the linear regime is observed. This indicates that the superconducting path resembles a thin line
but can not be fully considered as such.

The transition from $L=0 \rightarrow 1$ is strongly delayed when increasing the asymmetry. The first vortex enters the sample at $\Phi=1.3~\Phi_0$ for
the symmetric sample and at $\Phi=1.4~\Phi_0$ and $\Phi=2.9~\Phi_0$ for $a=0.3$ and $0.6~\mu m$ respectively. Baelus~{\it et al.\/}\cite{baelus00} indeed
found a delay of the penetration of the first vortex by increasing the displacement of the hole from the center but not with a factor larger than two as
seen for the sample with $a=0.6~\mu m$. The next transition ($L=1 \rightarrow 2$) is not clearly resolved. We could however see that the transition
occurs first for the sample with $a=0.3~\mu m$ at $\Phi \approx 3.4~\Phi_0$, followed by the symmetric ring at $\Phi=3.7~\Phi_0$, and then only at
$\Phi=4.7~\Phi_0$ for the strongly asymmetric sample, i.e. the sample with the largest off the center shift of the hole. The theoretical investigations,
however, showed a decreasing field for the entry of the second vortex by increasing the asymmetry. The calculations were not performed at the phase
boundary but deep in the superconducting state. Also a slightly smaller hole than in our experiments and different material parameters were used. This
could be a reason for the discrepancy for the sample with $a=0.6~\mu m$. Deep in the superconducting state, the complete structure will be in the
superconducting state, while at the measured phase boundary, only the region around the hole is superconducting so that a smaller effective area should
be taken, to explain the strong increase of the magnetic field value where the transitions between $L \leftrightarrow L+1$ take place in the sample with
the hole displaced over the largest distance from the center.

Baelus~{\it et al.\/}\cite{baelus00} found that at $L=1$ the vortex was trapped in the hole and that at $L=2$ and 3, the hole captures one
$\Phi_0$-vortex. A second vortex (2$\Phi_0$-vortex for $L=3$) is placed across the axis of displacement of the hole but at the opposite position. In our
experiment, we believe that the sample area where the second vortex was found in the calculation is not yet in the superconducting state. In this case,
no vortex can sit there and the hole can also not trap any vortex since no superconducting path around the hole exists. The vortices must then be placed
at a position between the contacts and the hole. This configuration is similar to the case of a infinite wedge where confined circulating supercurrents
were predicted in the vicinity of the corner\cite{klimin99}.

\section{Dissipation below $T_c(H)$}
\label{sec:diss}

\begin{figure*}
\centering
\includegraphics*[width=6cm,clip=]{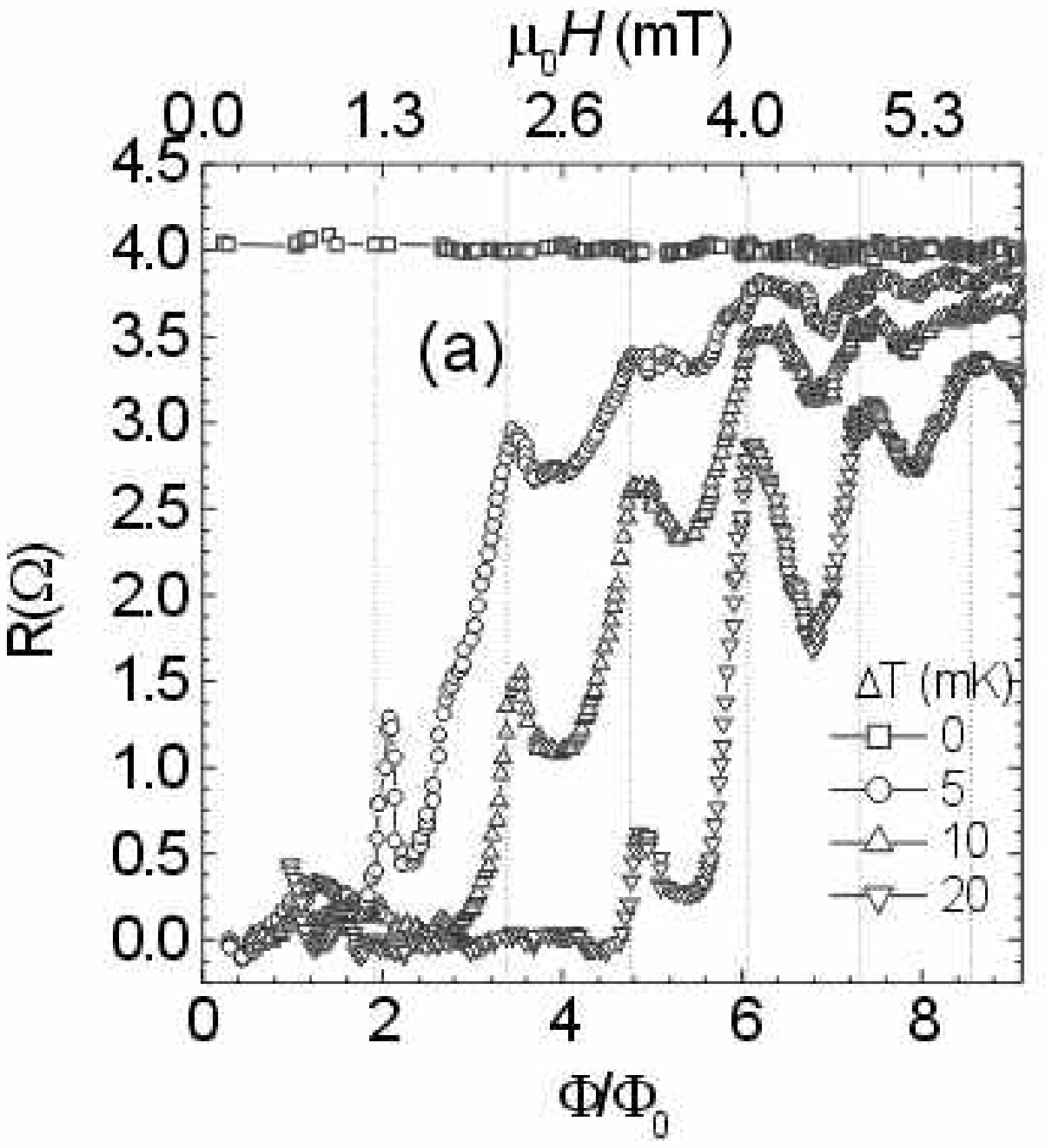}
\includegraphics*[width=6cm,clip=]{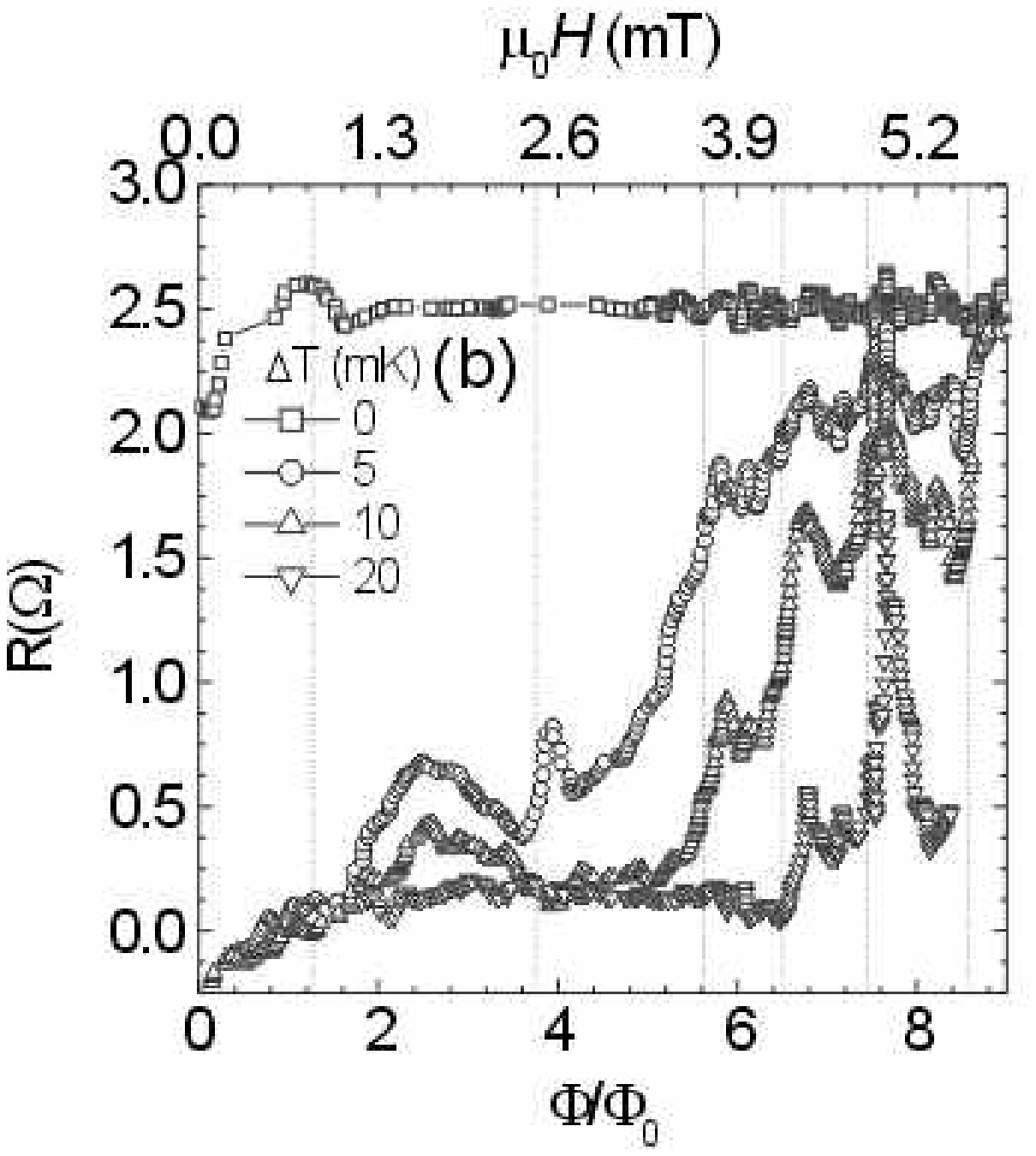}
\includegraphics*[width=6cm,clip=]{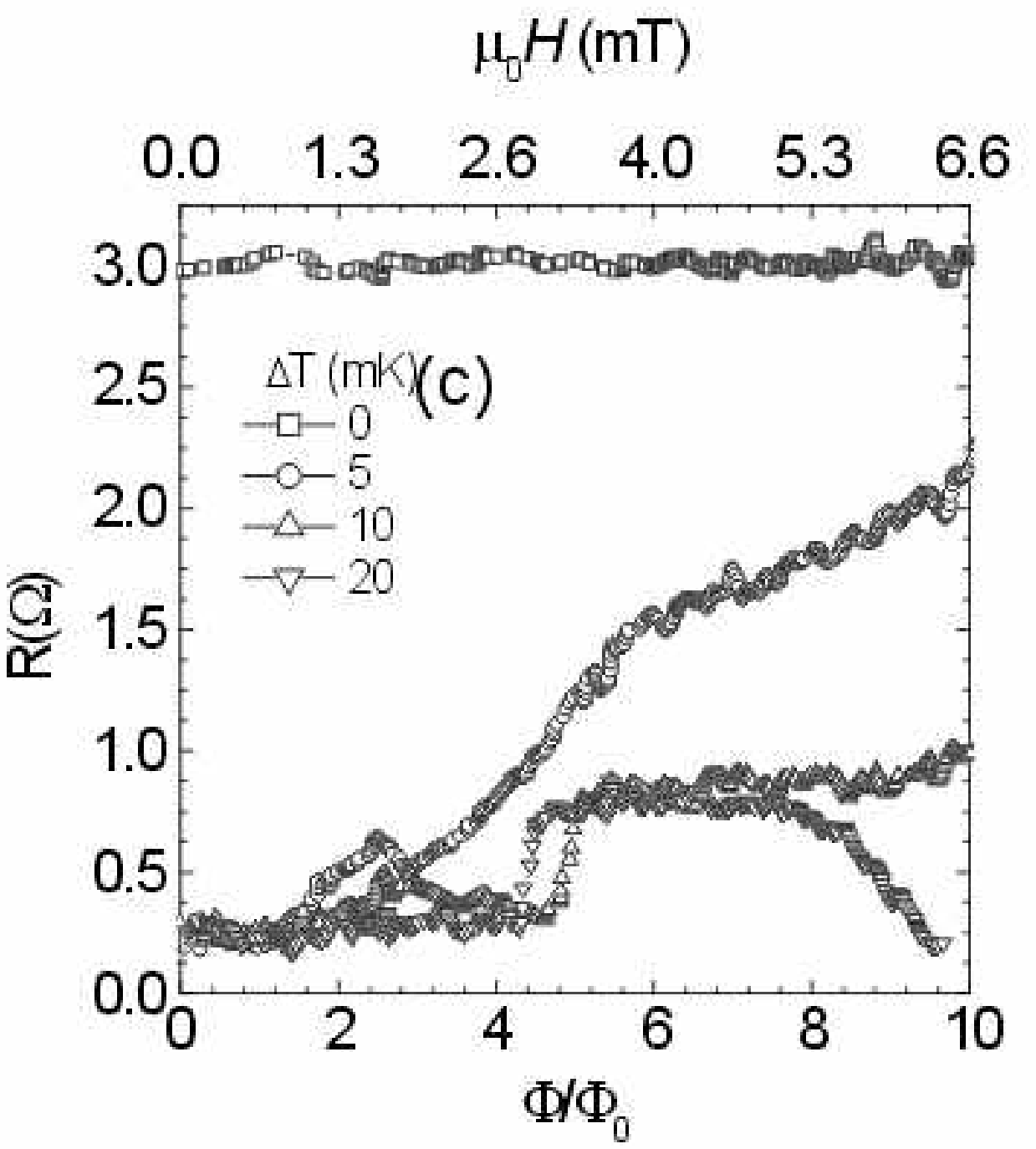}
\includegraphics*[width=6cm,clip=]{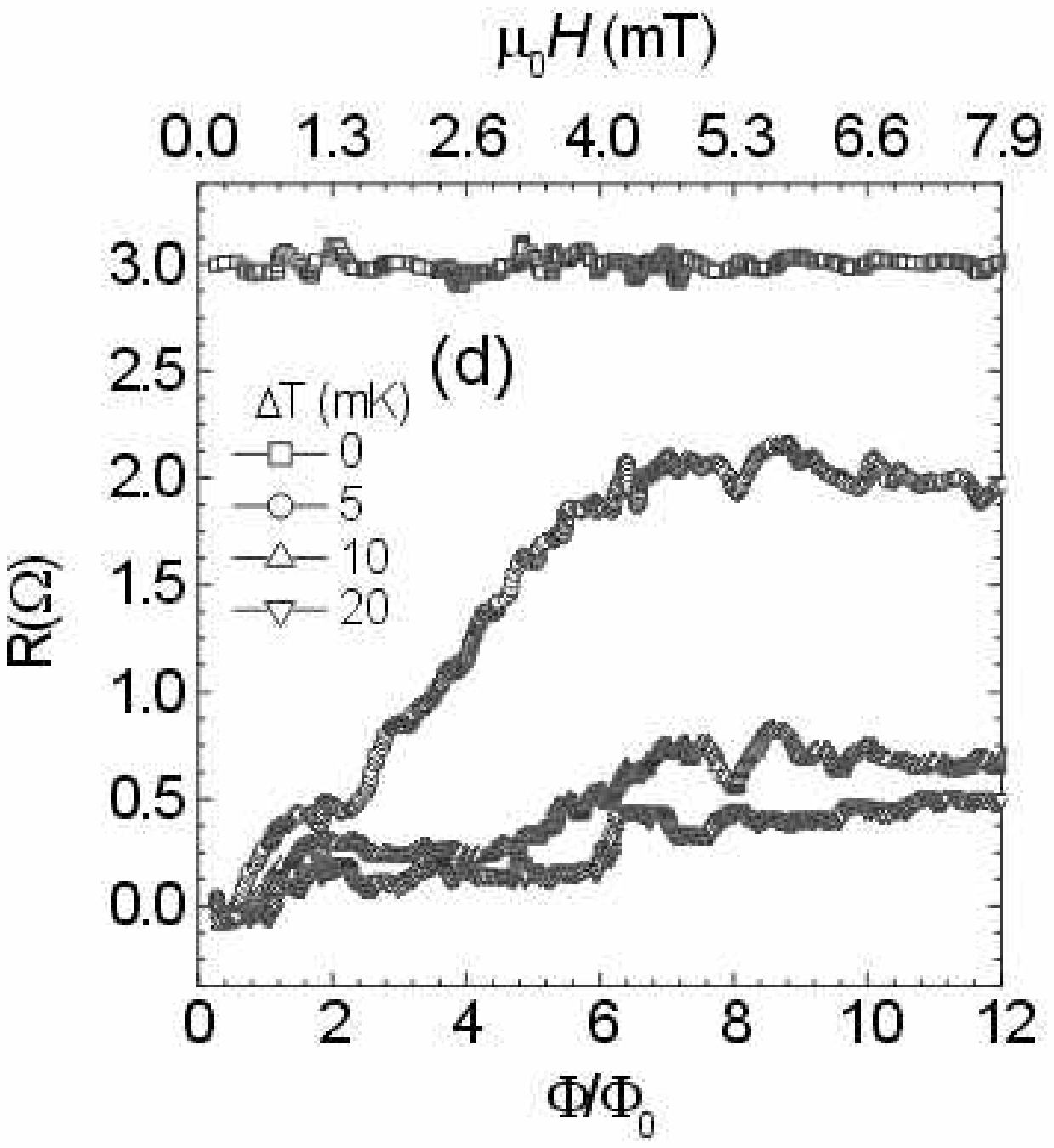}
\caption{Resistance of (a) the disk and of the ring with outer radius $r_o$=1~$\mu m$ and with a hole radius $r_i$=0.3~$\mu m$. The hole is (b) in the
center and  moved (c) by 0.3~$\mu m$ and (d) by 0.6~$\mu m$ from the center. The resistance is measured at a temperature $T(H)=T_c(H)-\Delta T$
corresponding to a shift of $\Delta T$ = 0, 5, 10, and 20 mK below the experimental phase boundaries.} \label{Fig:RBdecentr}
\end{figure*}

In order to analyze the onset of dissipation below $T_c(H)$, the resistance has been measured at a certain fixed temperature below the phase boundary.
Assuming that the resistance criterion ($R_c=2/3 R_n$) used for the determination of the phase diagram of the disk presented in Fig.~\ref{Fig:TcBLoop}(a)
is correct, the experimental phase boundary was shifted along the temperature axis with a value $\Delta T$ and the resistance has been measured following
the translated phase boundary, i.e. at a temperature equal to $T(H)=T_c(H)-\Delta T$. These curves were obtained from the same set of $R(H)$ curves as
for the determination of the phase boundary. The result for the disk is shown in Fig.~\ref{Fig:RBdecentr}(a). The curve with zero shift gives the
resistance criterion of $2/3~R_n$ as expected. For temperatures below the phase boundary, two different parts are directly distinguished. For low fields,
an almost zero resistance is measured, even for the smallest shift of the phase boundary. At higher fields, a resistive region is found. The region where
no resistance is observed corresponds to the magnetic field value where the vorticity is zero in the disk ($\mu_0H<1.3$ mT). Once that the first vortex
enters the sample a resistive behavior is noticed. The reason for the observation of a finite resistance can be found in the presence of the external
current used to measure the phase boundary. Just below the critical temperature of the disk, the value of the order parameter is very low so that a low
current will easily destroy locally superconductivity. When going deeper in the superconducting state, a larger area can sustain the applied current up
to a certain temperature where a superconducting path over the sample is found and a zero resistance is measured. This however can not fully explain the
magnetic field dependence shown in Fig.~\ref{Fig:RBdecentr}(a). Another possible dissipation mechanism, also related to the applied current, could be the
motion of the vortices. A current $\vec{I}$ will generate a Lorentz force $\vec{F}_L\propto\vec{I} \times \vec{\Phi}_0$ on the vortices. The electric
fields generated by the vortex motion can cause dissipation of energy that is characterized by the observation of a finite resistance in transport
measurements. The dissipation can also be caused by the nucleation of phase slips centers in the sample. It was found in~\cite{davidovic97prb} that the
change of vorticity in a superconducting loop transit through a phase slip state associated with a smaller energy barrier for the transition from $L$ to
$L \pm 1$.

Pronounced oscillations are present in the resistance curves of the disk. While in magnetoresistance curves at a constant temperature, the appearance of
oscillations is directly related to the presence of cusps in the $T_c(H)$ line. Here no oscillations should be expected since the resistance is measured
at a fixed temperature interval below the phase boundary. The position of the peaks in Fig.~\ref{Fig:RBdecentr}(a) corresponds to the magnetic fields
where cusps are observed in the phase boundary [dotted lines in Figs.~\ref{Fig:TcBLoop}(a) and \ref{Fig:RBdecentr}(a)]. The existence of the oscillations
suggests that the dissipation mechanism strongly depends on the stability of the vortices. The large amplitude of the oscillation shows that it is easy
to move the vortices only when the magnetic field is close to the value where the vorticity changes: $L \rightarrow L+1$. At these field values, a
constant fluctuation between the states with vorticity $L$ and $L+1$ will probably occur. Vortices will then enter from one side and will leave the
sample at the other side in the direction imposed by the Lorentz force. At high vorticity a large dissipation is observed indicating that the motion of
vortices is more pronounced when more vortices are present in the sample.

The resistance of the rings measured at a temperature $T(H)=T_c(H)-\Delta T$ below the experimental phase boundary is shown in Fig~\ref{Fig:RBdecentr}
for the rings with $a=0$, 0.3 and 0.6~$\mu m$, respectively. The dissipation in the circular symmetric ring resembles that of the disk. Peaks are seen at
the magnetic field value where the transition between two states with different vorticity takes place. The resistance at a fixed temperature interval
below the critical temperature seems to be smaller than for the disk without opening. The hole in the center of the disk will most probably act as an
artificial pinning center, preventing the vortices to move. The dissipation in the asymmetric samples starts to grow as soon as one vortex enters the
sample, as in the symmetric sample. However, for larger magnetic fields, the two asymmetric samples have a very different behavior. The oscillations seen
in the curves of  the symmetric sample are almost completely suppressed. Moreover, the resistance is not continuously growing but seems to saturate above
a certain magnetic field. For $\Delta T=20$~mK, the resistance even decreases with increasing magnetic fields for the sample with $a=0.3~\mu m$. The
dissipation is much lower than for the symmetric sample. All these observations indicate the presence of a quite different mechanism of the vortex motion
in the two asymmetric samples compared to the symmetric structures. Unfortunately, all the measurements were performed using an ac current. By comparing
the dissipation measured with dc current in two directions, it could be possible to detect if a preferential trajectory for the vortex motion exists.
Since the samples are not symmetric around the line between the two current contacts, the Bean-Livingston barrier should also be asymmetric. It is then
natural to expect a vortex motion that is dependent on the sign of the applied current, so called `vortex diode' effect\cite{villegas03}

\section{Conclusions}

We have studied the nucleation of superconductivity in doubly connected superconductors in the form of thin superconducting disks with a circular
opening. The effect of the size and of the position of the hole on the superconducting properties of the structures has been investigated. A parabolic
background of $T_c$ with periodic oscillations is found for the thinnest loops. For disks with smaller holes, a transition from a 1D regime to a 2D
regime is seen when increasing the magnetic field. For high magnetic field, the loops recover the behavior of the disk without opening. A giant vortex
state is then formed and the opening in the middle of the disk does not play an important role anymore.

The experimental results of the rings of different wire width were compared with theoretical calculations in the framework of the linearized GL equation.
Good agreement between our experimental results and the calculation of $T_c(H)$ were found. Small deviations in the amplitude of the oscillations were
observed. Moreover, for the thinnest loop that was studied, two different resistance criteria had to be used for low and high magnetic fields. These
deviations were explained by the fact that the theoretical linear phase boundary of the contacts  is crossing the parabolic $T_c(H)$ line of the thin
loops. At low magnetic fields, the nucleation occurs first in the loop while at higher magnetic fields superconductivity develops first in the wedge
shaped current and voltage contacts. As a consequence, a resistance criterion, dependent on the magnetic field, should be used for the determination of
$T_c(H)$.

Breaking the symmetry by moving the hole away from the center increases the critical field. The displaced hole forms a small region where
superconductivity is enhanced. A superconducting path for the applied current is likely to be formed before superconductivity nucleates in the whole
sample. The supercurrent cannot flow around the hole so that the singly connected state is recovered for a loop. The dissipation mechanism due to vortex
motion is strongly altered in this case.

It has been observed that the phase boundary is not affected by the presence and by the dimensions of a hole as long as no vortex is trapped inside the
sample. This, however, is only valid when the circular symmetry of the structure is kept. Once that the symmetry is broken by shifting the hole from the
center, the phase boundary at $L=0$ deviates from the $T_c(H)$ line of a disk.

\begin{acknowledgments}
This work has been supported by the Belgian IUAP, the Flemish FWO, the Research Fund K.U.Leuven GOA/2004/02 programmes and by the ESF programme VORTEX.
M.M. acknowledges support from the Institute for the Promotion of Innovation through Science and Technology in Flanders (IWT-Vlaanderen). The authors
would like to thank W.V. Pogosov for useful discussions and O. Popova for the X-rays measurements.
\end{acknowledgments}

\end{document}